\documentclass[preprintnumbers,amsmath,amssymb,nofootinbib,superscriptaddress]{revtex4}

\usepackage{amsmath}
\usepackage{amsfonts}
\usepackage{amssymb}
\usepackage{graphicx, rotating}
\usepackage{epstopdf}
\usepackage{epsfig}
\usepackage{latexsym}
\usepackage{graphicx}
\usepackage{color}
\usepackage{amsmath,amssymb}
\usepackage{slashed}
\usepackage{hyperref}
\hypersetup{colorlinks, citecolor=bluscuro, linkcolor=black, urlcolor=bluscuro}
\definecolor{rossos}{cmyk}{0,1,1,0.55}
\definecolor{bluscuro}{rgb}{0.15, 0.2, .85}
\definecolor{bluchiaro}{cmyk}{1,.3,0.,0.1}

\newcommand{\eq}[1]{Eq.~(\ref{#1})}

\newcommand{\be}{\begin{equation}}
\newcommand{\ee}{\end{equation}}
\newcommand{\beq}{\begin{equation}}
\newcommand{\eeq}{\end{equation}}
\newcommand{\bea}{\begin{eqnarray}}
\newcommand{\eea}{\end{eqnarray}}

\newcommand{\GeV}{\,\mathrm{GeV}}

\def\s{s_\beta}
\def\c{c_\beta}

\begin{document}

\title{SUSY Faces its Higgs Couplings}
\date{\today}

\author{Rick S. Gupta}
\email{sgupta@ifae.es}
\affiliation{IFAE, Universitat Aut{\`o}noma de Barcelona,
08193 Bellaterra, Barcelona, Spain}

\author{Marc Montull}
\email{mmontull@ifae.es}
\affiliation{IFAE, Universitat Aut{\`o}noma de Barcelona,
08193 Bellaterra, Barcelona, Spain}

\author{Francesco Riva}
\email{friva@ifae.es}
\affiliation{IFAE, Universitat Aut{\`o}noma de Barcelona,
08193 Bellaterra, Barcelona, Spain}
\affiliation{Institut de Th\'eorie des Ph\'enom\`enes Physiques, EPFL,1015 Lausanne, Switzerland}

\begin{abstract}
In supersymmetric models, a correlation exists between the structure of the Higgs sector quartic potential and the coupling of the lightest CP-even Higgs to fermions and gauge bosons.
We exploit this connection to relate the observed value of the Higgs mass $m_h\approx 125$ GeV to the magnitude of its couplings. We analyze different scenarios ranging from the MSSM with heavy stops to more natural models with additional non-decoupling D-term/F-term contributions. A comparison with the most recent LHC data, allows to extract bounds on the heavy Higgs boson masses, competitive with bounds from direct searches.
\end{abstract}

\maketitle

\noindent

\section{Motivation}

The quest for SUSY has taken an unexpected turn with the Higgs discovery at 125 GeV~\cite{ATLAS}.
Indeed, it is well known that the supersymmetric contribution to the Higgs mass is at most $(m_h^{tree})^2\lesssim m_Z^2$, implying that a large portion of the Higgs mass $\Delta m_h^2\gtrsim 86^2 \GeV^2$ must originate from symmetry breaking effects. Within the MSSM, for large stop masses, top/stop loops provide this necessary contribution, but only at the expense of naturalness, as the large loop effects needed to increase the Higgs mass also destabilize the EW scale. Experiments are therefore telling us that, if SUSY exists, it is either tuned, or it doesn't fulfill Occam's principle  and that more complicated models, with additional contributions to the Higgs quartic, have to be considered.

Still, a common feature of most SUSY models\footnote{An exception is the model of Ref.~\cite{Riva:2012hz} where the Higgs is the neutrino superpartner and there are no extra Higgs doublets. Ref.~\cite{Davies:2011mp} also proposes a model with one doublet only  while in Refs.~\cite{Alves:2012ez,Gupta:2009wn} additional doublets have been studied.}, is the Higgs sector, containing at least a particular version of a two Higgs doublets model (2HDM). Mixings in this extended Higgs sector,  lead to modified tree-level couplings between the lightest CP-even Higgs and the SM gauge bosons 
and fermions,
and provides a distinctive signature of SUSY, complementary to direct searches. While the latter remain the most favorable strategy for SUSY searches (in particular in the most natural SUSY realizations, where states associated with the stabilization of the electroweak (EW) scale are expected to be light), modified couplings could be the strongest evidence for SUSY in particular regions of parameter space, such as those with compressed spectra.

Interestingly, in 2HDMs, a correlation exists between the Higgs mass and its tree-level couplings to SM fields. Indeed, any  contribution to the Higgs quartic  potential, necessary in SUSY models to increase the Higgs mass from its tree-level value up to the observed value of  approximately 125 GeV, also changes the relation between  mass  and  hypercharge eigenstates and modifies the couplings of the  lightest CP-even Higgs. In this article we investigate this correlation in detail, showing how different models that accommodate the observed Higgs mass also modify their Higgs couplings. We then confront these expectations with the most recent LHC data \cite{ATLAS:2012tx019}-\cite{ATLASPhotons8}, which we use to extract limits on the parameter space of such theories (in particular on $m_A$ and $\tan\beta$).

We first show, with a simple and intuitive analytical approximation, how Higgs mass and couplings are correlated in SUSY models (or 2HDMs in general) (section~\ref{sec:mass-couplings}). Then we study, in turn, the MSSM with heavy stops (section~\ref{sec:MSSMstops}), the MSSM with extra non-decoupling D-terms  (section~\ref{sec:DTerms}), and the F-term contributions of NMSSM-like models (section~\ref{sec:BMSSM}), where we also discuss a general class of models beyond the MSSM (BMSSM). In section~\ref{sec:lightstops} we comment on how these conclusions are modified in the presence of sizable loop-effects due to light SUSY partners and we leave for Appendix I the details related to our global fits and for Appendix II a summary of the formulas used in our plots.

\section{The Higgs Mass/Couplings Connection} \label{sec:mass-couplings}\label{sec:connection}
\begin{figure}[h]
\begin{center}
\includegraphics[width=5cm]{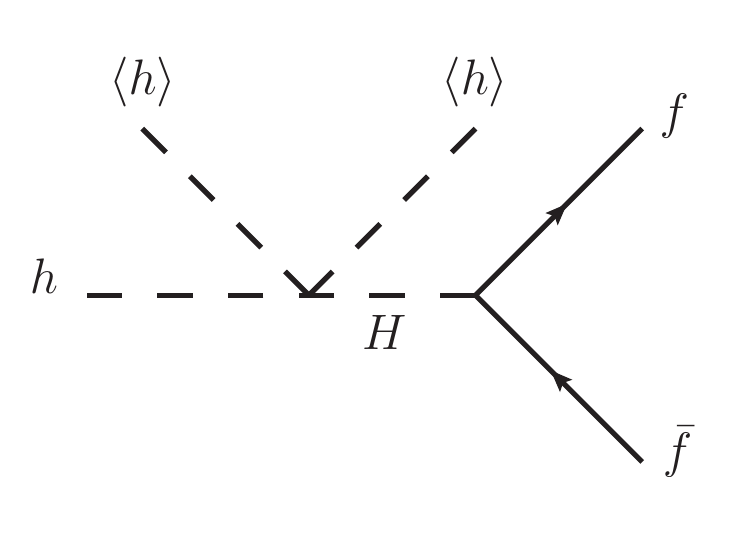}
\end{center}
     \caption{ \emph{The mixing between $h$ and $H$, induced by the quartic interaction $\delta h^3 H$, modifies the couplings of $h$ to the fermions w.r.t to its SM value.}}\label{fig:feyndiag1}
\end{figure}

Supersymmetry requires the existence of two Higgs doublets, $H_{1,2}$ giving mass to leptons and down-type/up-type quarks. Limiting our discussion to the third family fermions, which have the strongest couplings to the Higgs sector, we consider
\begin{equation}\label{superpot}
\mathcal{L}\supset -Y_b H_1 \bar q b - Y_t H_2 \bar q t -Y_\tau H_1 \bar l \tau\, .
\end{equation}
Only a linear combination of $H_1$ and $H_2$ obtains a vacuum expectation value (vev) $v\equiv 174 \GeV$; its couplings to SM fermions and vectors equal those of a SM Higgs. Any quartic contribution to the scalar potential for $H_1$ and $H_2$ introduces, in general, a misalignment between this linear combination and the mass eigenstates: this misalignment is responsible for a modification in the Higgs couplings. The best way to see this is in the basis $h,H$, where only one state ($h$) has a vev. The angle
 $\beta$ denotes the angle between these states and the neutral CP-even components of the  gauge eigenstates $H_1,H_2$:
\bea\label{rotation}
h_1^0&=&\cos\beta  h+ \sin\beta H\\
h_2^0&=&\sin\beta h-\cos\beta  H.\nonumber
\eea
In this basis, the couplings \eq{superpot} of $h$ and $H$ to fermions are,
\beq\label{YBYT}
-\cos \beta Y_b(h+\tan{\beta }H)\bar{b}b,\quad -\sin \beta	Y_t(h-\cot{\beta }H)\bar{t}t ,
\eeq
where couplings to charged leptons have the same form as for down-type quarks.
Now, consider a general contribution to the quartic of the Higgs potential written in terms of $h,H$,
\begin{equation}\label{effl}
\Delta V(H_1,H_2)= +\delta_\lambda h^4+\delta h^3 H+
\delta_2 h^2 H^2+\delta_3 h H^3+\delta_4 H^4,
\end{equation}
where the $\delta$'s are  given dimensionless couplings. The first term contributes to the lightest CP-even Higgs mass as
\beq\label{deltamh}
\Delta m_h^2 =16 \delta_\lambda v^2;
\eeq
in order to account for the observed value $m_h^2\approx 125\GeV$,
\begin{equation}
\Delta m_h^2= m_h^{obs\, 2}-m_Z^2 (\cos 2\beta)^2 \gtrsim (86\GeV)^2
\end{equation}
is needed. Interestingly, the same physics that is responsible for $\delta_\lambda$, also generates a mixing between $h$ and $H$, via the term $\delta h^3 H$, that leads to a modification of the Higgs couplings, as illustrated in FIG.~\ref{fig:feyndiag1}.
We can quantify these modifications, in the limit where $H$ is heavy, by integrating out the heaviest eigenstate from the relevant part of the Lagrangian
\begin{equation}\label{LH}
\mathcal{L}\supset -\delta h^3 H  -\sum_{f=t,b,\tau} Y^H_f\bar{f}f H-\frac{m_H^2}{2}H^2,
\end{equation}
where $Y^H_f$, the coupling of $H$ to fermion $f=t,b,\tau$, can be read from \eq{YBYT}. For large $m_H$ we can solve the equations of motion of $H$, giving $H\approx \delta h^3 Y^H_f\bar{f}f/m_H^2$,  and obtain the effective interaction
\begin{equation}\label{leff}
\mathcal{L}_{eff}\supset \delta  \sum_{f=t,b,\tau}Y^H_f h\bar{f}f\frac{h^2}{m_H^2}.
\end{equation}
Now, notice that the equations of motion for $H$ imply a small vev $\langle H\rangle \approx 2\sqrt{2} \delta (v^3/m_H^2)$, so that the expression for the fermion mass is modified accordingly and we can write the coupling of the physical Higgs $\tilde{h}=h-\sqrt{2}\,v$, normalized with its SM value $y_f^{SM}=m_f/v$, as
\begin{equation}
c_f\equiv \frac{y_{f}}{m_f/v}\approx\frac{Y_f^h- 6Y_f^H \delta \frac{v^2}{m_H^2}}{Y_f^h- 2Y_f^H \delta \frac{v^2}{m_H^2}}\approx 1-4\delta \frac{Y_f^H}{Y_f^h}  \frac{v^2}{m_H^2}\, .
\end{equation}
Using \eq{YBYT} to read $Y^{h,H}_f$, we finally obtain
\bea\label{candb}
c_{b,\tau}&\approx1-4\tan{\beta } \delta \frac{v^2}{m_H^2}\, ,\nonumber\\
c_t&\approx1+4\cot{\beta } \delta \frac{v^2}{m_H^2}\, .
\eea
This simple, yet important, expression summarizes the goal of this work: any new physics that is responsible for the large Higgs mass \eq{deltamh} also affects the Higgs couplings to fermions. This approximate formula allows us to understand qualitatively how this connection works and predicts whether a given contribution to the Higgs mass results in an increase or decrease of the couplings to tops and bottoms/taus (similar methods have been used in Refs.~\cite{Blum:2012kn,Azatov:2012wq,Blum:2012ii} to study Higgs couplings modifications). Nevertheless, notice that in our plots we always use the exact expressions listed in Appendix~II,Ê rather than \eq{candb}.

Deviations in the Higgs couplings to vectors can be studied in a similar way, giving
\begin{equation}\label{vectorcoupl}
c_{V}= 1-\mathcal{O}\left(\delta^2\frac{v^4}{m_H^4}\right)
\end{equation}
which is  generally suppressed w.r.t. deviations in the couplings to fermions (we have checked that in the region preferred by data this statement holds at better then the 2 \% level and deviations in $c_V$ can be ignored).

In principle, complete analyses of Higgs couplings in a SUSY context should take into account possible
modifications of the tree-level couplings to up-type quarks, to down-type quarks (and leptons) and to vectors; at the loop level extra contributions from light SUSY partners to the couplings to gluons and photons could be present, and in total generality also the possibility of an invisible decay width should be considered (see Ref.~\cite{Riva:2012hz} for a motivated scenario were the Higgs can decay invisibly in a SUSY context): a total of six parameters (see Refs.~\cite{FITS,azatovgall} for a list of recent analyses of this type). Nevertheless, ignoring the last possibility, \eq{vectorcoupl} tells us that in the simplest SUSY models, couplings to vectors are not expected to deviate much from the SM ones (this is not true when the Higgs sector is extended to include extra states in different $SU(2)_L$ representations that can mix with the Higgs, as we shall discuss in section~\ref{withmix}). Furthermore, the null results of direct SUSY searches suggest that SUSY partners should have masses of a few hundreds GeV and that their loop contributions to the effective $hgg$ and $h\gamma\gamma$ couplings might be small (we comment about this  in section~\ref{sec:lightstops}). For these reasons, in what follows, we orient our analysis mostly to the  Higgs couplings to tops and to bottoms/taus and compare theoretical expectations with data through an intuitive simplified scenario where only $c_t$,$c_b$ are free to vary, and all other couplings are fixed to their SM values.


 \section{The Minimal Supersymmetric Standard Model}\label{sec:MSSMstops}

The technique of the previous section can be applied also to  the tree-level contribution of the Minimal Supersymmetric Standard Model (MSSM)\footnote{In this case, $h$ and $H$ can be thought  of as the eigenstate of the mass matrix before electroweak symmetry breaking.}. The only contribution to the quartic potential comes from the D-term which, for  the $SU(2)_L\times U(1)_Y$  MSSM gauge group, reads
\begin{equation}\label{VMSSM}
\Delta V_{MSSM}=\frac{g^{2}+g^{\prime\,2}}{8}\left(|H_1^0|^2-|H_2^0|^2\right)^2=\frac{g^{2}+g^{\prime\,2}}{32}\left((c_{\beta }^2-s_{\beta }^2)^2 h^4+8(c_{\beta }^2-s_{\beta }^2)s_{\beta }c_{\beta } h^3 H+\cdots\right)
\end{equation}
with $c_{\beta }\equiv \cos\beta $ and $s_{\beta }\equiv \sin\beta $ and in what follows we shall also use $t_\beta\equiv\tan\beta$. This defines
\bea\label{delta1}
\delta_\lambda&=&\frac{m_Z^2}{16 v^2}(\c^2-\s^2 )^2,\\
\delta&=&\frac{m_Z^2}{ 2v^2} \s \c (\c^2-\s^2).
\label{delta2}
\eea
From \eq{candb}, this gives
\bea
c_b&\approx& 1-\frac{m_Z^2}{2m_H^2}\sin 4 \beta  \tan \beta\\
c_t&\approx& 1+ \frac{m_Z^2}{2 m_H^2}\ \sin 4 \beta \cot \beta. 
\label{MSSMapprox}
\eea
which coincides with the usual decoupling limit of the MSSM~Ê\cite{Djouadirev} with the identification $m_H\approx m_A$ (which is accurate for $m_{A,H}\gg m_Z$ or in the large $\tan\beta$ limit), and we will use in what follows in the comparison between exact and approximate results. At the same time, \eq{delta1} provides the well known  contribution to the Higgs mass $m_h^2=m_Z^2 \cos^2 2 \beta$; this tree-level result is modified by loop effects, in particular from top quarks/squarks, which we consider in what follows.

\subsection{Top Squarks with no mixing}\label{sec:heavystop}

\begin{figure}
\begin{center}
\begin{tabular}{cc}
\includegraphics[width=0.5\columnwidth]{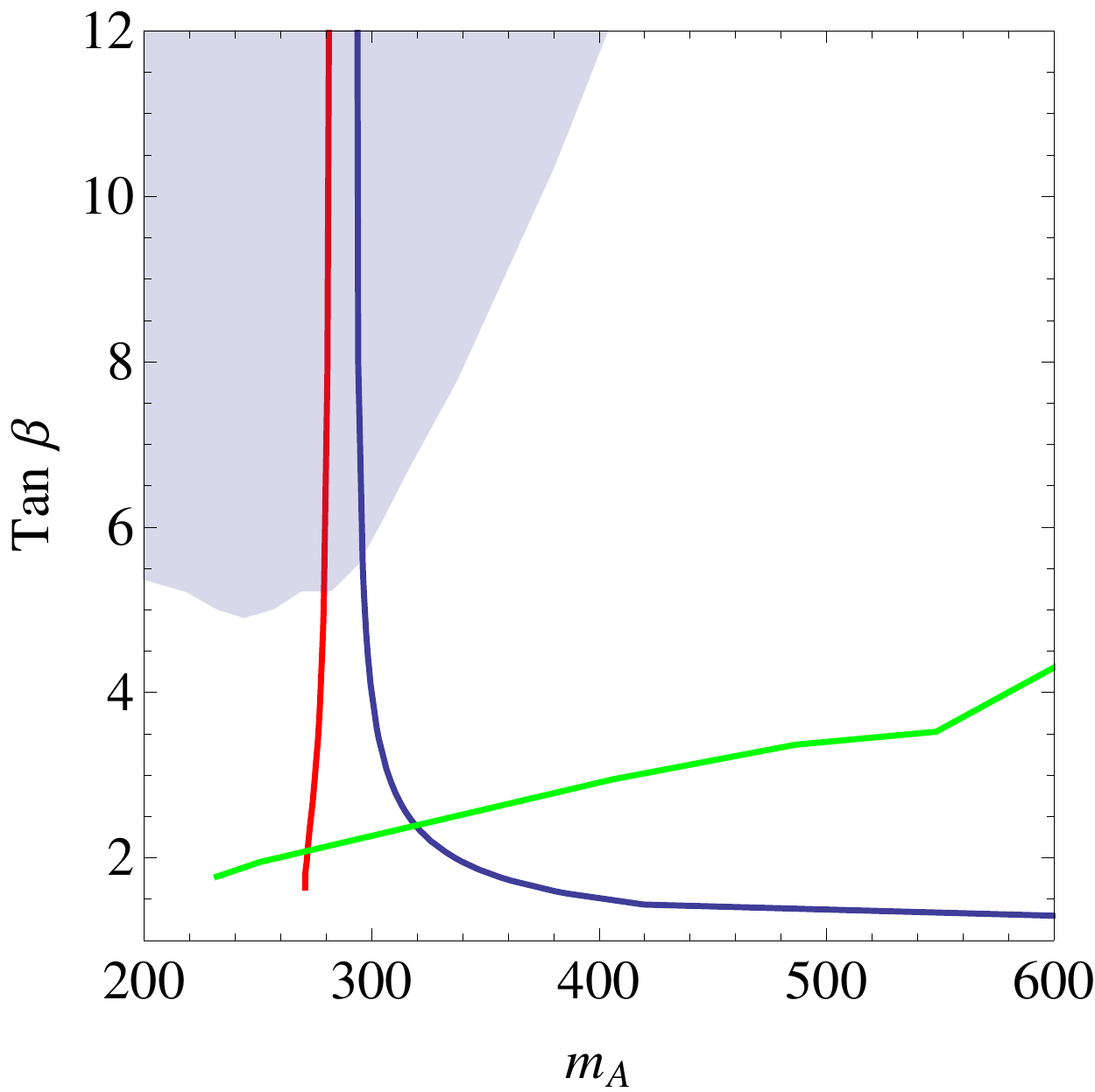}&
\includegraphics[width=0.5\columnwidth]{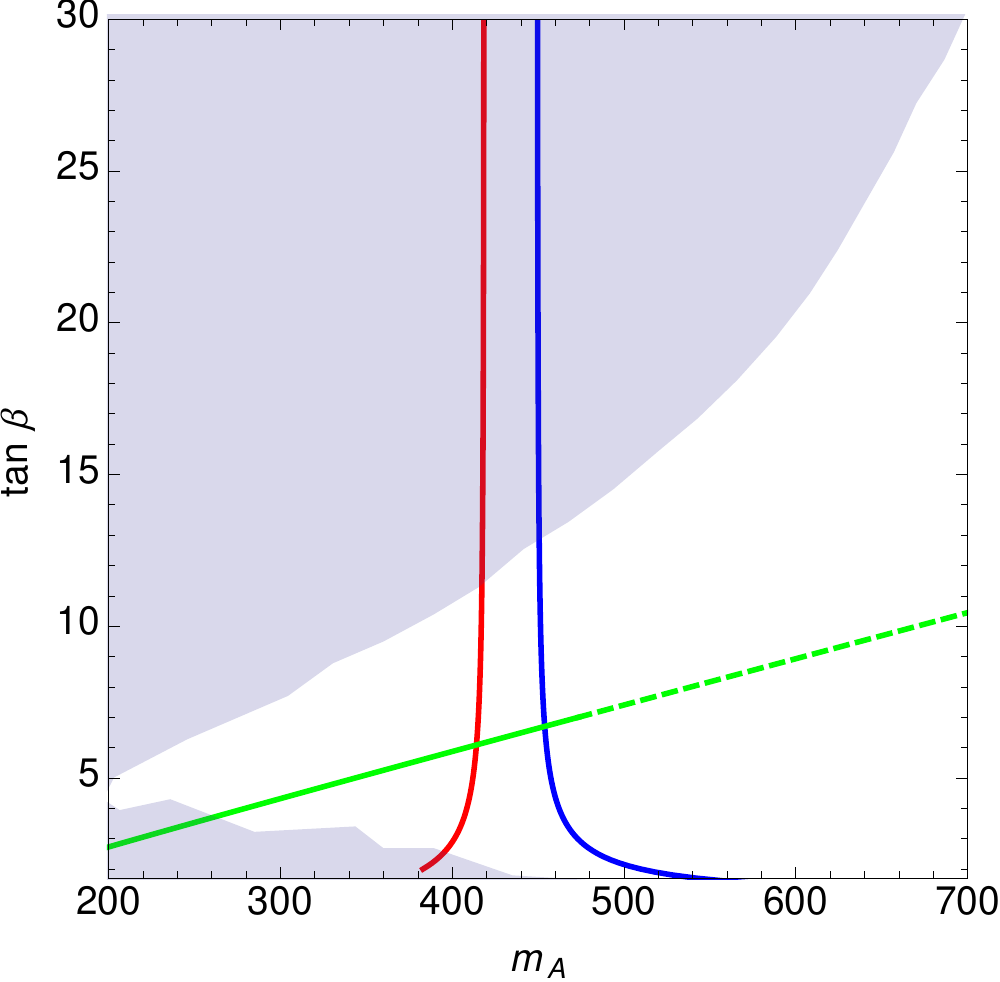}
\end{tabular}
\end{center}
\caption{\emph{Exclusion plot in the $m_A,\tan\beta$ plane for the MSSM with heavy stops (red), for models with additional non-decoupling D-terms (blue) and F-terms (green); regions to the left of the lines are excluded.  The shaded region corresponds to bounds from direct searches \cite{CMSHtautau}. Left: present data; right: longterm projection based on \cite{Peskin:2012we} assuming no deviations from the SM, shaded region from Ref.~\cite{Linssen:2012hp} (the dashed part of the line corresponds to a region where $\lambda_S$ is bigger than 2 and non reaches the non-perturbative regime below approximately 10 TeV \cite{Lodone:2010kt,Hall:2011aa}).}}
\label{fig:excl}
\end{figure}

We begin with the case of top squarks with no mixing (realized in popular SUSY breaking mechanisms such as gauge mediation and gaugino mediation where  a small trilinear coupling is expected~\cite{Maloney:2004rc}). The dominant loop contribution to the scalar effective  potential is \cite{Djouadirev,Carena:1995bx},
\beq\label{deltaVstops}
\Delta V_{stop}=\frac{\lambda_{2} }{2} |H_2|^4\, ,
\eeq
where,
\beq\label{lambdastops}
\lambda_2 \approx \frac{3 y_t ^4}{8 \pi^2} \log[m_{\tilde{t}_1}m_{\tilde{t}_2}/M^2_t]
\eeq
 (a more accurate expression can be found in Appendix II). After rotating into the basis of \eq{rotation} one identifies
\bea
\label{deltastops1}
\delta_\lambda&=& s_\beta^4 \frac{\lambda_{2}} {8}\\
\delta&=&-4 s_\beta^3 c_\beta \frac{\lambda_{2}} {8}. 
\label{deltastops2}
\eea
From  \eq{lambdastops} and from \eq{deltastops1} it follows that, in order to obtain a Higgs mass compatible with experiment, multi-TeV stop masses are required. Such heavy stops also destabilize the EW scale through loop effects and push the MSSM into fine-tuning territory~\cite{Hall:2011aa}. Ignoring for a moment this tension, we can assume these loop contributions to be uniquely responsible for the large value of the Higgs mass, and write the deviations of $c_{b,t}$ induced by loop effects \eq{deltastops2} together with the ones from the tree-level potential \eq{delta2}, as
\bea
c_b &\approx&
1+ \frac{m_h^2-m_Z^2 \cos 2\beta }{m_H^2}\, , \nonumber\\
c_t &\approx&
1- (\cot \beta)^2  \frac{m_h^2-m_Z^2 \cos 2\beta }{m_H^2}.
\label{stopapprox}
\eea
This shows that, in the MSSM with no stops mixing and for $\tan\beta>1$, the deviations in $c_b$ ($c_t$) are always  positive (negaitive), as already observed in Ref.~\cite{Azatov:2012wq}. For large  $ \tan \beta$ the deviations in $c_t$ are suppressed, while
\begin{equation}\label{cbapproxMSSM}
(c_b-1) \approx \left(\frac{154\GeV}{m_H}\right)^2.
\end{equation}
\begin{figure}
\begin{center}
\includegraphics[width=0.6\columnwidth]{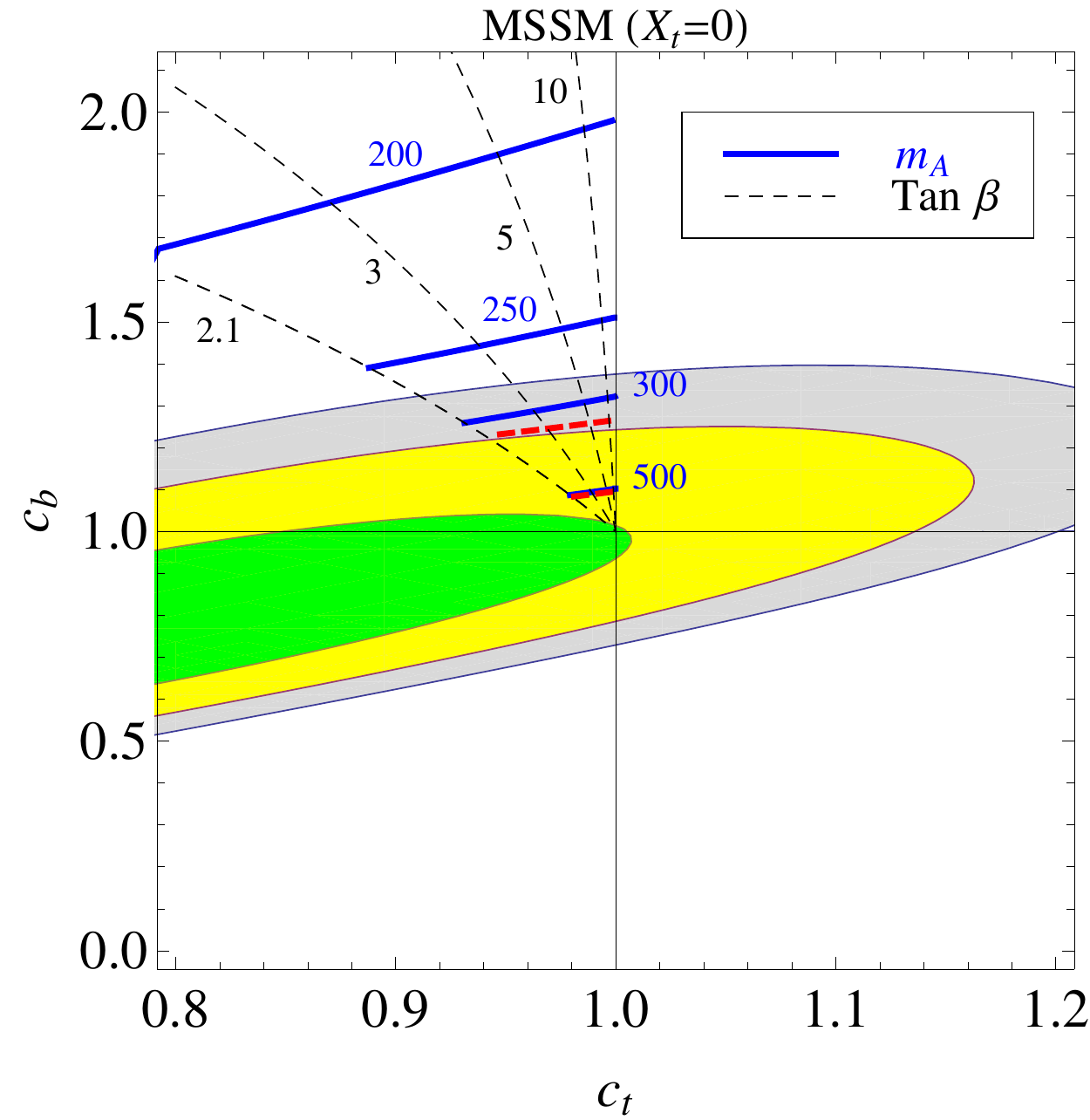}
\end{center}
\caption{\emph{Theoretical expectation for Higgs couplings deviations for the MSSM with heavy stops and no mixing, taking $m_h=125\GeV$, showing contours of constant $m_A$ (solid blue) and $\tan \beta$ (dashed), obtained from the exact expressions of Eqs.~(\ref{hmass},\ref{alpha}) of Appendix II. Also shown are the 68\% (green), 95\%(yellow) and 99\%(grey) C.L. regions obtained by a global fit of the most recent LHC Higgs data, as explained in Appendix I, neglecting loop contributions to the $hgg$ and $h\gamma\gamma$ couplings. The dashed red lines show the approximate results of  \eq{stopapprox} for $m_H=300,500\GeV$. }}
\label{stop}
\end{figure}
We can compare these results with the exact ones of Fig.~\ref{stop}, which shows the intuitive $(c_b,c_t)$-plane mentioned above, and compares these theoretical expectations with the most recent data \cite{ATLAS:2012tx019}-\cite{ATLASPhotons8}, using the methods described in Appendix I.  We assume a heavy sparticle spectrum, that does not affect the Higgs couplings to gluons and photons, other than through \eq{stopapprox} (this is motivated by the fact  that in this example, we are assuming multi-TeV stops).  Masses $m_H\lesssim 250 \GeV$ can be excluded, almost independently of $\tan\beta$, as suggested already by \eq{cbapproxMSSM} for a sensitivity to the $h\bar b b$ coupling of about 50\%. In Fig.~\ref{fig:excl} we also show the CMS bounds on the traditional MSSM $m_A,\tan \beta$ plane (for a recent analysis see Ref.~\cite{Arbey:2012dq}) from direct searches of the heavy Higgs decaying into $\tau$ pairs, as performed by CMS \cite{CMSHtautau}. As can be appreciated, analyses of the light Higgs couplings offer a complementary search strategy in the intermediate $\tan\beta$ region.

\subsection{Top Squarks with  mixing}

In the presence of sizable  A-terms, L and R top squarks can mix, inducing additional contributions to the Higgs effective potential~\cite{Carena:1995bx,Haber:1993an},
\beq
\Delta V^{mix} = \frac{\lambda_2}{2} |H_2|^4+ ( \frac{\lambda_5} {2}|H_1H_2|^2 +\lambda_7  |H_2|^2 H_1H_2+c.c),
\eeq
where the values of $\lambda_2$, $\lambda_5$ and $\lambda_7$ depend in particular on the parameter $\mu$ and the trilinear $A_t$  and their expression, at the one loop level, can be found in Appendix II. In the point of `maximal mixing', when the trilinear term is $|A_t-\mu\cot\beta|=\sqrt{6} m_{\tilde{t}}$ (where $m_{\tilde{t}}$ is the geometric mean of the lightest stop masses), the contribution to the Higgs mass proportional to  $\lambda_2$ is maximized, while $\lambda_7=0$. Recasting the potential in the $h,H$ basis gives,
\bea\label{maximal1}
\delta_\lambda &=& s_\beta^4\left(\frac{\lambda_2}{8}  + \frac{\lambda_5}{4\, t_\beta^2}   +\frac{ \lambda_7 }{2\,t_\beta}\right)\, ,\\
\delta&=&s_\beta^3 c_\beta\left( \frac{\lambda_2}{2}  + \frac{\lambda_5}{2} \left(1-\frac{1}{t_\beta^2}\right) +\frac{ \lambda_7 }{2}\frac{t_\beta^2-3}{\sqrt{t_\beta^2+1}}\right)\, ,
\label{maximal2}
\eea
where it can be seen that for large $\tan\beta$ (which is necessary in the MSSM to maximize the tree-level mass), the dominant contribution to the Higgs mass still comes from the first term $\lambda_2$, similarly to the case with no mixing discussed in the previous paragraph. As mentioned above, this term is maximized by large mixing, with drastic effects and the stop mass can be as low as  550 GeV in this case. Nevertheless, a fine-tuning at the percent level persists  due to the fact that large $A_t$ terms also contribute to the Higgs mass-parameter~\cite{Hall:2011aa}.

\begin{figure}
\begin{center}
\begin{tabular}{cc}
\includegraphics[width=0.45\columnwidth]{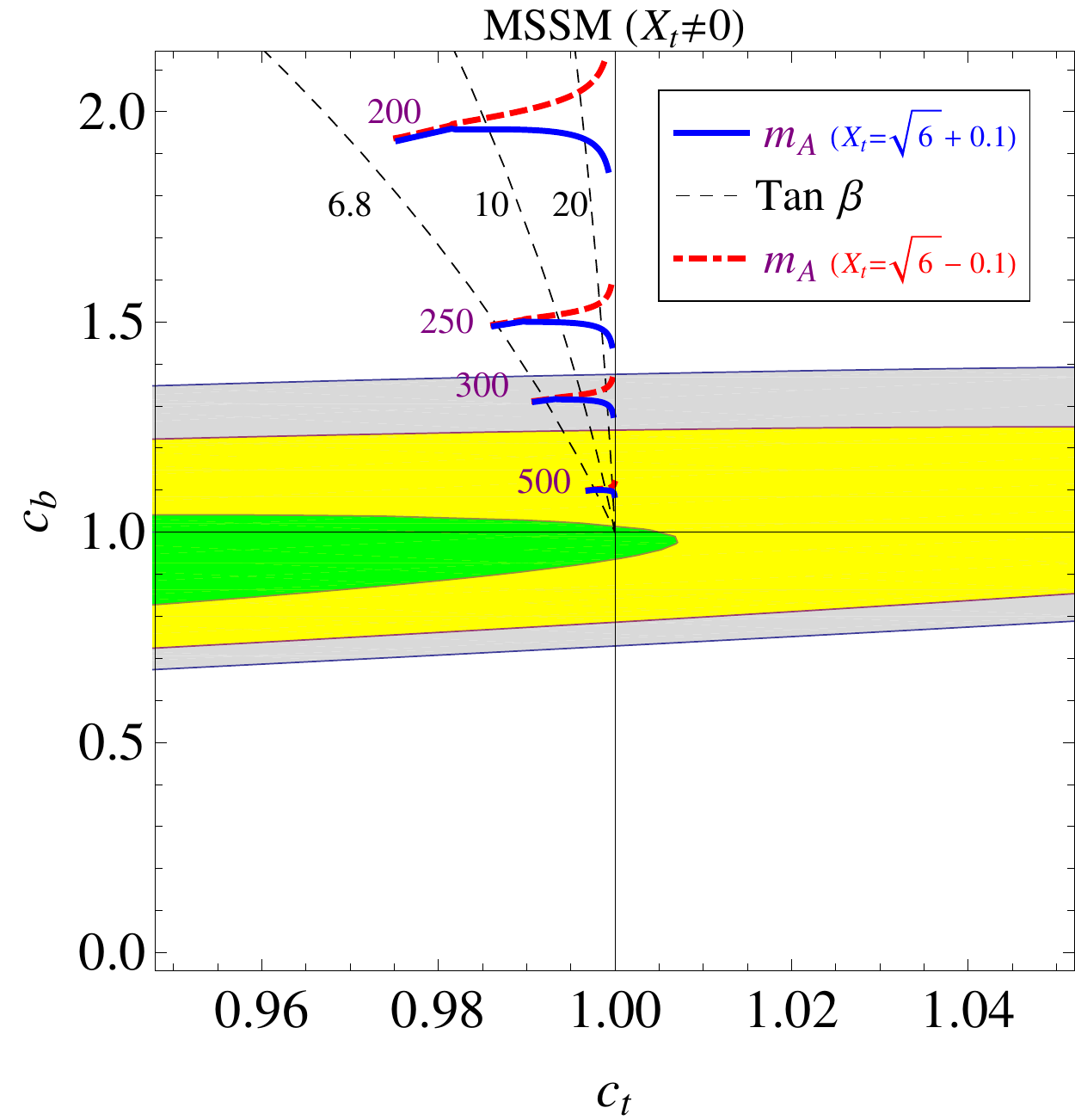}
\end{tabular}
\end{center}
	\caption{\emph{Same as FIG.\ref{stop}, but for near maximal mixing and, again, we adjust $\sqrt{m_{\tilde{t}_1}m_{\tilde{t}_2}}\in [550, 2000]$ GeV in order to obtain the observed Higgs mass. We take $x_t = \sqrt{6}\pm0.1$ for the blue/red curve in order to show the influence, for large $\tan\beta$, of small deviations from maximal mixing; $\mu=400 \GeV$.}}\label{stop2}
\end{figure}

Unfortunately, for a generic choice of $\mu$ and $A_t$, the multitude of parameters introduced by mixing weakens the Higgs mass/coupling connection as shown by \eq{maximal2} where sizable $\lambda_{5,7}$ can affect the Higgs couplings without contributing to the Higgs mass. We show this effect in Fig.~\ref{stop2} where we consider small deviations from maximal mixing:   departures from $\lambda_7=\lambda_7^{MaxMix}=0$ are enhanced at large $\tan\beta\gtrsim 20$ and the contribution to $\delta$ and to our predictions can be seizable. Nevertheless such large values of $\tan\beta$ are already in tension with rare $B$ processes, such as $B_s\to \mu^+\mu^-$ \cite{Altmannshofer:2012ks}, and with direct searches for $H/A\to \bar\tau \tau$ \cite{CMSHtautau}, so that we do not expect our results to change significantly in the intermediate $\tan\beta$ region, where our bounds are more competitive, see Fig.~\ref{fig:excl}.

\section{Extra D-Terms}\label{sec:DTerms}

As discussed above, a 125 GeV Higgs in the MSSM is generally associated with fine-tuning. This suggests that the principle of SUSY, if realized at low energy in a natural way, extends beyond the MSSM, with new tree-level effects contributing to the Higgs quartic. The first possibility is to envisage additional gauge symmetries that contribute to the Higgs quartic, similarly to the MSSM gauge group~\cite{Maloney:2004rc, Batra:2003nj,Lodone:2010kt}.  In this section we study the example of an additional abelian gauge group under which $H_1$ and $H_2$ have opposite charges (as compatible with the $\mu$-term). Then, the extra contribution to the Higgs sector quartic\footnote{The form of the potential in \eq{dpot} holds also for the non-abelian extension considered in Refs~\cite{Batra:2003nj,Lodone:2010kt}. }
\begin{equation}
\Delta V=\kappa \left(|H_1^0|^2-|H_2^0|^2\right)^2
\label{dpot}
\end{equation}
 where,
\beq\label{kappaetc}
\kappa= \frac{g_X^2}{8(1+\frac{M_{Z'}^2}{2 m_\phi^2})}.
\eeq
%
%
Here $m_\phi$ is the soft SUSY breaking mass of the MSSM singlets that breaks the  U(1)$_X$ group (with gauge coupling $g_X$) and $M_{Z'}$ the SUSY-preserving mass of the gauge boson. Eq.~(\ref{kappaetc}) shows that, in the limit  $M_{Z'} \gg m_\phi$, the $Z'$ can be supersymmetrically integrated out and the D-term contribution of the  U(1)$_X$ group decouples: non-decoupling D-terms require a large soft mass $ m_\phi\sim M_{Z'} $ and result in an effective hard breaking in the Higgs sector.

\begin{figure}
\begin{center}
\includegraphics[width=0.6\columnwidth]{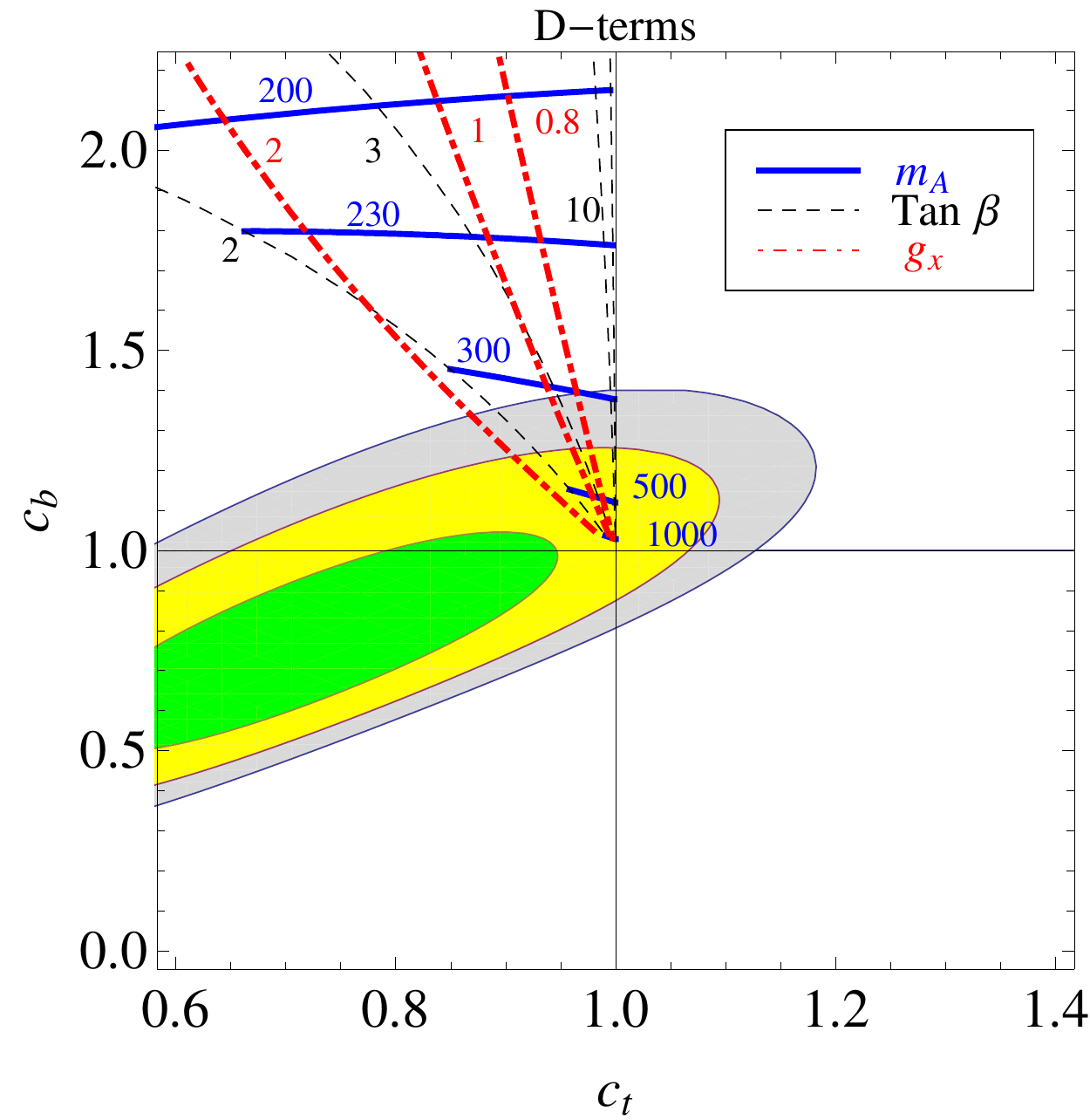}
\end{center}
\caption{\emph{Higgs couplings deviations in the MSSM with additional non-decoupling D-terms  to raise the Higgs mass to 125 GeV (on top of the effect of light stops $m_{\tilde{t}}=500\GeV$). The global fit (Colors as in Fig.~\ref{stop}) includes the effect of a 500 GeV stop (to be compared with Fig.~\ref{stop} where the effects of stops on the fit are vanishingly small).} }
\label{dterm}
\end{figure}

The contributions to $\delta_\lambda$ and $\delta$ are similar to Eqs.~(\ref{delta1},\ref{delta2}), with the substitution $m_Z^2/v^2\to4\kappa$. In the absence of other effects that affect the Higgs mass (we assume the loop effects of Eqs.~(\ref{deltastops2},\ref{maximal1}) to be subdominant), we can fix $\kappa$ in order to obtain the observed Higgs mass \footnote{Notice that as $\tan \beta \to 1$, all contributions to the Higgs mass from D-terms vanish; hence these expressions have to be trusted only away from this singular point: in FIG.~\ref{dterm} we show curves of constant $g_X$ (in the limit of large $m_\phi\gg M_{Z^\prime}$) to show that in the region of interest the parameters are under control.}, we can then write
\bea \label{eol}
c_b&\approx &1+2\frac{m_h^2}{m_H^2} \frac{t_{\beta }^2}{t_{\beta }^2-1} \\
c_t &\approx &1-2\frac{m_h^2}{m_H^2}\frac{1}{t_{\beta }^2-1}.
\eea
meaning that, for $\tan\beta>1$, positive (negative) deviations are expected in $c_b$ ($c_t$). For large $\tan\beta$ the modifications in $c_t$ vanish, as usual, while those on $c_b$ asymptote to $c_b-1\approx (176 \GeV/m_H)^2$. This is shown, using the exact expressions from Appendix II, in Fig.~\ref{dterm}. Differently from Fig.~\ref{stop}, the global fit of Fig.~\ref{dterm} includes the effect of a light stop at $500$ GeV (as opposed to the previous section, where heavy stops were necessary to increase the Higgs mass, here this is taken care by the additional D-terms, and the stops can be naturally light, see also Section~\ref{sec:lightstops}).  Masses $m_H\lesssim 300\GeV$ can already be excluded, with better results in the small $\tan\beta$ region (see also Fig.~\ref{fig:excl}).

In principle we could relax the assumption that $H_1$ and $H_2$ carry equal and opposite $U(1)_X$ charges. In this case, however, additional structure is needed in order to generate a $\mu$-term. For example an extra SM singlet, charged under $U(1)_X$ can generate this term by aquiring  a non-vanishing  vev. This extension, however, implies additional contributions to the quartic potential from F-terms which, as we comment in the next-section, are expected to dominate.


\section{F-Terms, the NMSSM and the BMSSM }\label{sec:BMSSM}

It is tempting to parametrize these new effects using an effective field theory approach with an expansion in powers of the scale of physics beyond the MSSM (in the example of the previous section, this would be the mass of the new gauge bosons $M_{Z^\prime}$). The most general such parametrization, however, lacks any predictive power (peculiar directions in parameter space can be found where an increase in the Higgs quartic coupling doesn't imply modifications of the couplings \cite{CraigNew}). Nevertheless, as shown in Ref.~\cite{Dine:2007xi}, the leading order effects in such an expansion have a very specific form\footnote{For large $\tan\beta$ interactions at higher order in the expansion could be enhanced and dominate.}:
\beq\label{LBMSSM}
{\cal L}_5= \int d^2 \theta \left(\frac{\lambda_1}{M} (H_1 H_2)^2 + {\cal Z}\frac{\lambda_2}{M} (H_1 H_2)^2 \right)
\eeq
where ${\cal Z}= \theta^2 m_{SUSY}$ is a dimensionless spurion that parametrizes SUSY breaking. This leads to additional contributions to the scalar potential,
\beq
\Delta V_{5}=2  \epsilon_1 H_1 H_2 (H_1^\dagger H_1+H_2^\dagger H_2) + \epsilon_2 (H_1 H_2)^2 +c.c
\eeq
with $\epsilon_1 = \lambda_1 \mu^*/M$ and $\epsilon_2 = -\lambda_2 m_{SUSY}/M$. We obtain
\bea
\delta_\lambda&=& \frac{\epsilon_1}{4} \sin 2 \beta+ \frac{ \epsilon_2}{16} \sin^2 2 \beta\nonumber\\
\delta&=& - \frac{\epsilon_1}{2} \cos 2 \beta- \frac{ \epsilon_2}{8} \sin 4 \beta.
\eea
By construction $\beta\in[0,\pi/2]$ and for the first term to contribute positively to the Higgs mass, a positive $\epsilon_1$ is necessary, implying an enhancement of $c_b$ and a decrease in $c_t$, similarly to the case studied in the previous section. The term proportional to $\epsilon_2$, on the other hand, reduces for $\tan\beta>1$ the $h\bar bb$ coupling while increasing the coupling to top quarks, oppositely to the effects of D-terms. This is an interesting case that corresponds to the non-decoupling F-term contribution of an extra singlet, interacting with the Higgs sector via the superpotential term $W=\lambda_S S H_1 H_2$, as in the NMSSM. Indeed, in the limit where the mass of the singlet is large, its contribution is given by the second term of \eq{LBMSSM}, where $M=M_S$ ($m_{SUSY}=m_S$) is the supersymmetric (SUSY breaking) mass of the singlet, and $\lambda_2=\lambda_S^2$ (notice that the singlet also gives a generally subdominant contribution  to the first term of \eq{LBMSSM} with $\epsilon_1=-\mu^*\lambda_S^2/(2M_S)$, which we ignore for the time being).

\begin{figure}
\begin{center}
\includegraphics[width=0.5\columnwidth]{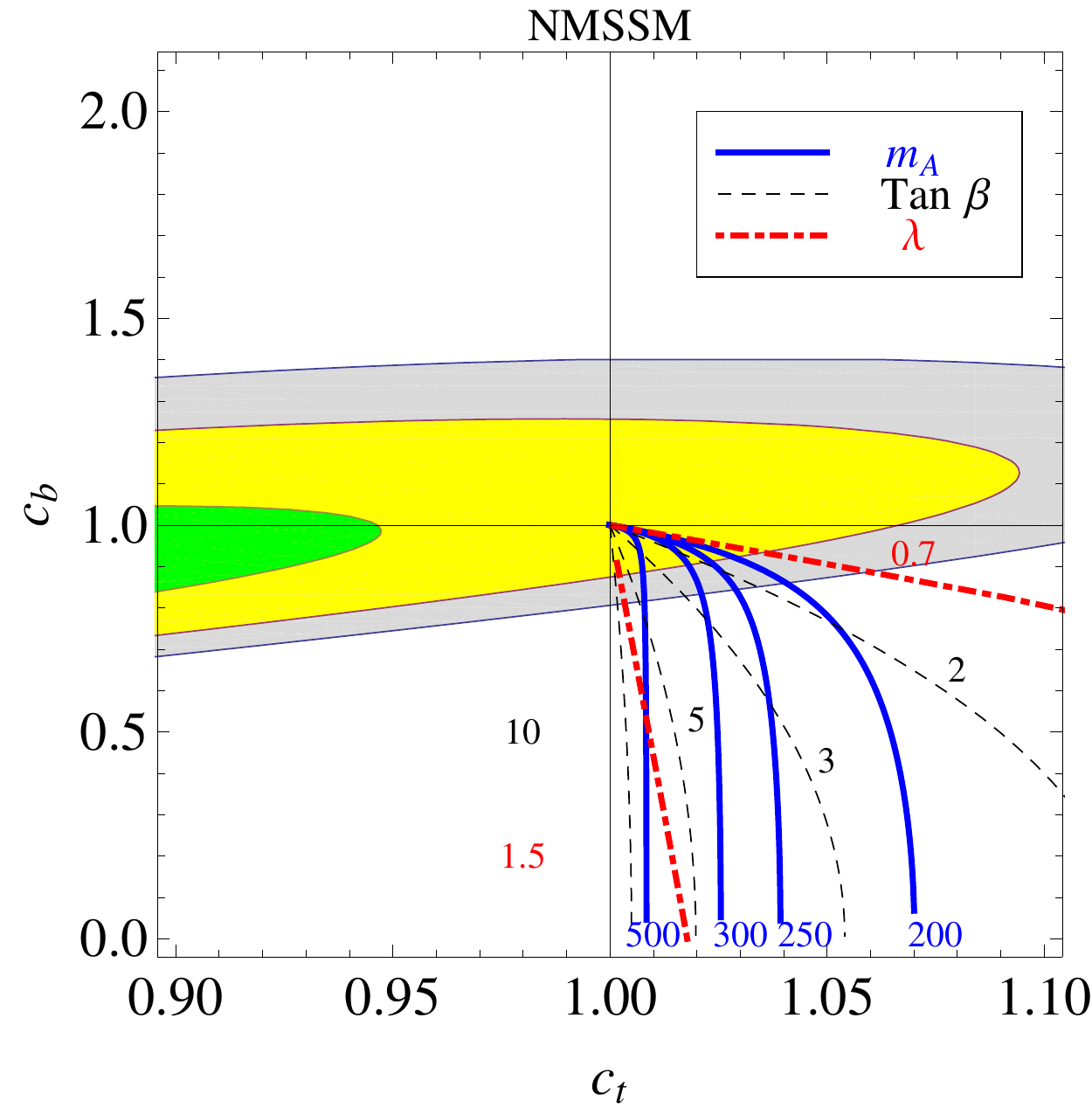}
\end{center}
\caption{\emph{Coupling deviations in the NMSSM assuming a  Higgs mass of 125 GeV in the limit where the singlet is heavy and it doesn't mix with the Higgs, but its contributions do not decouple. Global fit as in Fig.~\ref{dterm}.}}
\label{nmssm}
\end{figure}

If the largeness of the Higgs mass is due to a combination of the MSSM D-terms effects of \eq{VMSSM} and the present  contribution from F-terms due to the singlet (i.e. with negligible contributions from loop-effects), then the Higgs couplings to fermions are modified as
\bea
c_b&\approx& 1-\frac{t_\beta^2-1}{2}\frac{m_h^2-m_Z^2}{m_H^2}\\
c_t&\approx&1+\frac{t_\beta^2-1}{2t_\beta^2}\frac{m_h^2-m_Z^2}{m_H^2},
\eea
which, for large $\tan\beta$, gives deviations in the $h\bar tt$ coupling of order $\Delta c_t\approx(60 \GeV/m_H)^2$, and in the couplings to bottom quarks $\Delta c_b\approx t_\beta^2 (60 \GeV/m_H)^2$. We show  the exact coupling deviations in Fig.~\ref{nmssm} (we assume, again, the presence of  500 GeV stops, see section \ref{sec:lightstops}) where we also emphasize curves of constant $\lambda_S$: values below $\lambda_S\lesssim 0.7$ are perturbative up to the GUT scale, while for values $0.7\lesssim \lambda_S\lesssim 2$ the non-perturbative regime is reached above a scale of 10 TeV \cite{Barbieri:2007tu,Hall:2011aa}. The bounds on $m_H$ that can be extracted from this analysis are very much dependent on $\tan\beta$, as can be seen in Fig.~\ref{fig:excl}.

%
%

While the approach of \eq{LBMSSM} encompasses large classes of models, its applicability is limited to cases with widely separated scales, such as the NMSSM where the singlet has both a large SUSY preserving and SUSY breaking mass\footnote{
Triplets with hypercharge  $Y \pm 1$ and superpotential $W= \lambda_{T} T H_2 H_2 + \lambda_{\bar{T}}\bar{T} H_1 H_1$ have also been considered in the literature: in the non-decoupling limit, their contribution to the potential is
\beq
\Delta V= |\lambda_{T}|^2 H_2^4 +| \lambda_{\bar{T}}|^2H_1^4
\eeq
and
\beq
\delta_\lambda = \frac{| \lambda_{\bar{T}}|^2}{4}  c_\beta^4 + \frac{ |\lambda_{T}|^2 }{4}  s_\beta^4,\quad \delta= | \lambda_{\bar{T}}|^2  c_\beta^3 s_\beta+ |\lambda_{T}|^2  s_\beta^3 c_\beta.
 \eeq
For large $\tan \beta$ only the $H_2^4 $  term is important and the results coincide with those of section~Ê\ref{sec:heavystop}.
}. In the opposite case, however, its interactions with the Higgs sector can induce mixings with the lightest CP-even Higgs and the analysis changes completely, as we now discuss.

\subsection{Doublet-singlet mixing}\label{withmix}

\begin{figure}[!t!]
  \begin{minipage}{0.4\textwidth}
   \centering
   \includegraphics[scale=0.65]{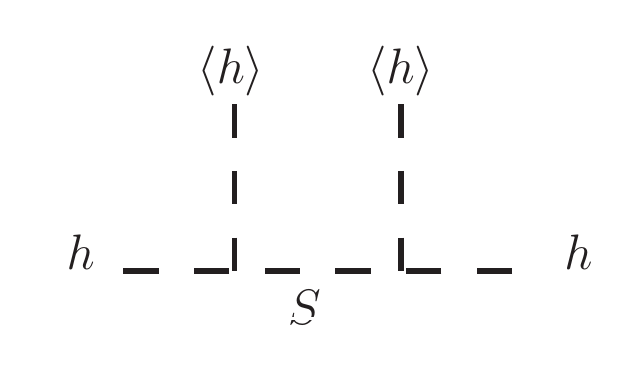}
    \end{minipage}\hspace{1.5 cm}
   \begin{minipage}{0.4\textwidth}
    \centering
    \includegraphics[scale=0.65]{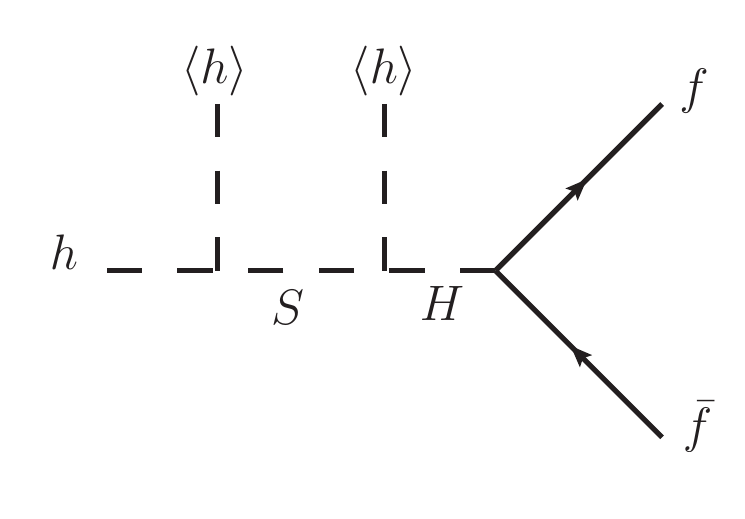}
    \end{minipage}
 \caption{\emph{Feynman diagrams illustrating modifications in the couplings between the Higgs and the fermions (or vectors) due to mixings with the singlet $S$.}}     \label{fig:feyndiag2}
 \label{fig:BoltzmannRelic}
\end{figure}

When the singlet is not much heavier than the  EW scale, the above analysis ceases to be valid; moreover singlet-Higgs mixing can affect our discussion of section~\ref{sec:connection} (see also Ref.~\cite{Bertolini:2012gu} for other LHC bounds on this possibility). Indeed, in this case, the potential \eq{effl} includes in particular
the term $\Delta V(H_1,H_2,S)\supset \delta_S s h^2/2$. Once $h$ gets a vev, this term leads to a mixing between $h$ and $S$ so that $h$ becomes a linear combination of the three gauge eigenstates: \eq{rotation} now must include
\beq
h= \cos \theta(\sin \beta  h_2^0- \cos \beta  h_1^0) + \sin \theta s.
\label{linear}
\eeq
The mixing $\theta$ can be estimated by using the techniques of section~\ref{sec:connection}: the term $(\delta_S/2) h^2 s$ corrects the $h$-propagator when $s$ is integrated out and two of the $h$ legs are replaced by  vevs, as illustrated in the first diagram of FIG.\ref{fig:feyndiag2}.
This correction, beside modifying the quartic structure $\delta$ and $\delta_\lambda$ as discussed so far, it also  universally affects all $h$ couplings by shifting the kinetic term to
\beq
\mathcal{L}_{eff}\supset(1+2\frac{\delta_S^2 v^2}{m_S^4})\frac{1}{2} \partial_\mu h \partial^\mu h\,\,.
\eeq
Indeed, making this kinetic term canonical leads to a universal  suppression of all $h$ couplings by the factor
\beq
 \cos \theta\approx1- \frac{\delta_S^2 v^2}{m_S^4} \,\,,
 \eeq
where $\theta$ is defined by \eq{linear} and the coupling  of Eqs.~(\ref{candb},\ref{vectorcoupl}) become,
\bea\label{modMSSMcv}
c_b&\approx&1-4\tan{\beta } \delta \frac{v^2}{m_H^2}-\frac{\delta_S^2 v^2}{m_S^4}\\
c_t&\approx&1+4\cot{\beta } \delta \frac{v^2}{m_H^2}- \frac{\delta_S^2 v^2}{m_S^4}\\
c_V&\approx& 1- \frac{\delta_S^2 v^2}{m_S^4} .\label{modMSSMcv3}
\eea
Notice that in this case, if the singlet is light or if its couplings to the Higgs sector are large, sizable modifications of the $hZZ$ and $hWW$ vertices can be produced. In principle,  it is still possible to exploit the Higgs mass/coupling connection   to fix $\delta$ and then a simultaneous measurement of $c_V$ and $c_{b,t}$ would allow to extract information about $m_H$ and about the mixing with the singlet. 
In practice, however, models of this type introduce many new contributions to the Higgs quartic potential and the Higgs mass/coupling connection looses most of its predictive power.
We show this in the example of the NMSSM~\cite{Ellis:1988er}, where the superpotential $W=\lambda SH_1H_2+\kappa S^3/3$ generates the following relevant terms in the potential (which add to the usual MSSM D-terms \eq{VMSSM}),
\beq
\Delta V_{NMSSM}(H_1,H_2,S) \supset m_S^2 S^2+ \lambda^2 \left(|H_1 H_2|^2 +S^2|H_1|^2+S^2|H_2|^2\right)-(\lambda A_\lambda S H_1H_2  + \lambda\kappa H_1H_2S^{*2}+h.c.)
\eeq
where we assume real coefficients for simplicity ($m_S,A_\lambda$ are soft SUSY breaking terms~\cite{Ellis:1988er}). After the singlet obtains a vev $\langle S\rangle\equiv v_S$, we can integrate out its real part, with mass $m_S$, and obtain the effective quartic potential
\beq
\Delta V_{NMSSM}^{eff}(H_1,H_2) \supset \lambda^2 |H_1 H_2|^2 -\frac{\mu_1^2}{m_S^2} \textrm{Re}(H_1H_2)^2-\frac{\mu_2^2}{ 4 m_S^2} (|H_1|^2+|H_2|^2)^2+\frac{ \mu_1 \mu_2}{m_S^2}\textrm{Re}(H_1H_2)(|H_1|^2+|H_2|^2),
\eeq
(where we have neglected higher order terms in the couplings) and the mixing term
\beq
\delta_S= \frac{(\mu_2 -\mu_1\sin 2 \beta ) }{\sqrt{2}},
\eeq
with $\mu_1 \equiv \lambda A_\lambda+2 \lambda\,  \kappa \,v_S $, and $\mu_2 \equiv 2  \lambda ^2  v_S$. In this procedure, also contributions from the second diagram of Fig.~\ref{fig:BoltzmannRelic} are taken into account. As usual the quartic potential can be written in terms of $h,H$ and we find,
\bea\label{deltaNMSSMmix1}
\delta_\lambda&=&\frac{\lambda^2}{16}\sin^2 2 \beta -\frac{1}{8}\frac{\delta_S^2}{m_S^2} \\
\delta&=&-\frac{\lambda^2}{8}\sin 4 \beta-\frac{ \mu_1 \delta_S}{ m_S^2} \frac{\cos 2\beta}{2\sqrt{2}}\label{deltaNMSSMmix2}
\eea

As it could have been foreseen, the multitude of parameters that characterize this model breaks the connection between $\delta$ and $\delta_\lambda$ and it becomes possible to raise the Higgs mass independently of a modification of its couplings. Even for small $m_H$ a conspiracy between the MSSM D-term  and these additional F-terms could allow for a large Higgs mass without any observable effect in the Higgs couplings (a similar example in the context of D-terms is discussed in Ref.~\cite{CraigNew}).

\begin{figure}
\begin{center}
\begin{tabular}{cc}
\includegraphics[width=0.5\columnwidth]{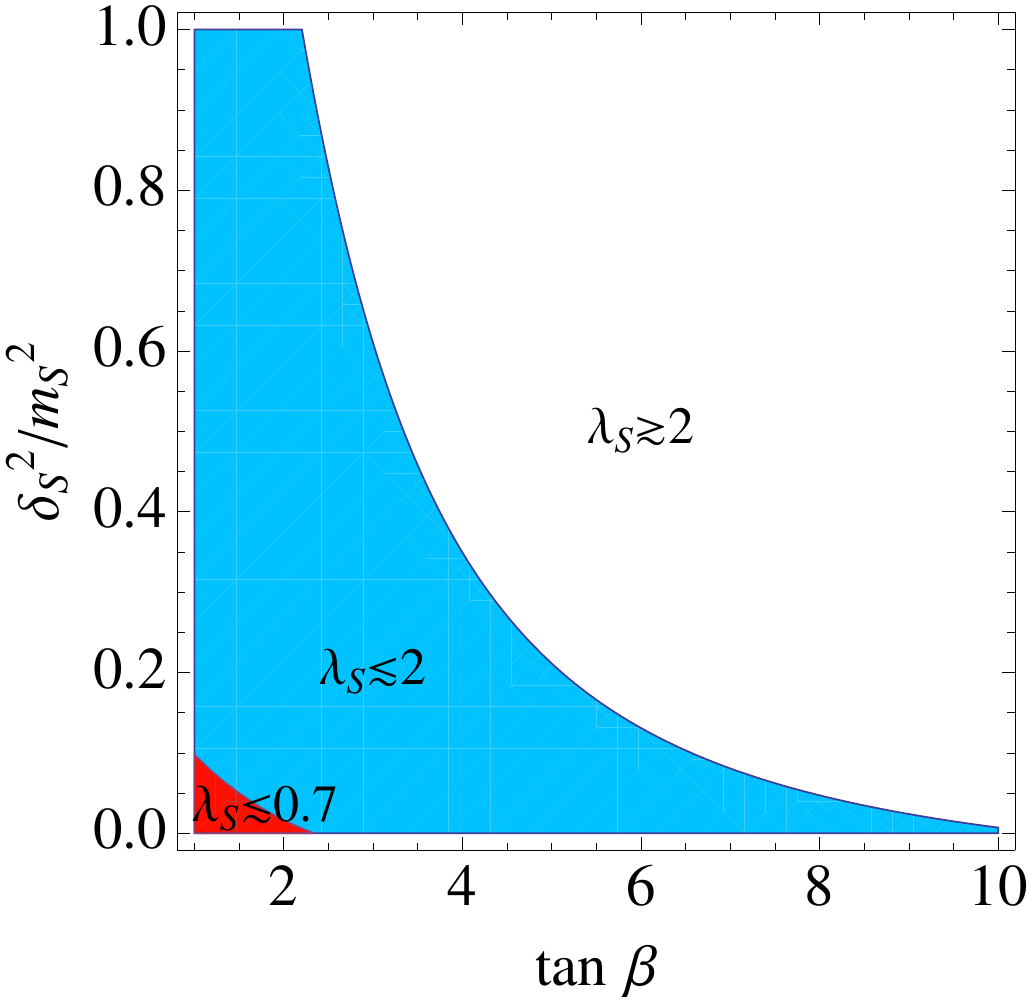}
\end{tabular}
\end{center}
\caption{\emph{Upper bound on $\delta_S^2/m_S^2$ as a function of $\tan\beta$ requiring perturbativity of the coupling $\lambda_S$ up to the GUT scale ($\lambda_S\lesssim 0.7$) or only up to 10 TeV ($\lambda_S\lesssim 2$), as in $\lambda$SUSY; the contribution of a 500 GeV stop is also included. The parameter $\delta_S^2/m_S^2$ enters the modifications of couplings to fermions and gauge bosons Eqs.~(\ref{modMSSMcv}-\ref{modMSSMcv3}) via the combination~$\delta_S^2 v^2/m_S^4$. }}
\label{nmssmix}
\end{figure}

Nevertheless, perturbativity up to the GUT scale (up to 10 TeV) limits the size of $\lambda\lesssim 0.7(2)$ and the necessity of a positive contribution to the Higgs mass from \eq{deltaNMSSMmix1}, imposes an upper bound on the negative contribution proportional to $\delta_S^2/m_S^2$, as we show in the left panel of Fig.~\ref{nmssmix}. Since the latter governs the coupling modification due to mixing through Eqs.~(\ref{modMSSMcv}-\ref{modMSSMcv3}), we see that in the perturbative NMSSM only small deviations are expected due to mixing, $\Delta c_V\lesssim 5\%$ for $m_S\gtrsim v$. Deviations in the couplings $c_{b,t}$ are still proportional to the parameter $\delta$ which, as mentioned above, is now independent of the Higgs mass and would allow only to constrain the ratio $\mu_1/m_H$, which is not particularly interesting.

In $\lambda$SUSY\cite{Barbieri:2006bg}, on the other hand, deviations can easily be of order unity. In particular, if $\delta$ in \eq{deltaNMSSMmix2} is positive (notice that for $\tan\beta>1$, both $\sin4\beta$ and $\cos2\beta$ are negative) we have $c_t\gtrsim c_V > c_b$, which enhances the rate of both $h\to\gamma\gamma$ and $h\to VV$. Notice that if we consider only deviations in the tree-level couplings, an enhancement of $h\to\gamma\gamma$ only, would require $c_V\gtrsim c_t \gtrsim c_b$, a region which is not touched by this model.

\section{Light SUSY Partners}\label{sec:lightstops}

\begin{figure}
\begin{center}
\begin{tabular}{cc}
\includegraphics[width=0.5\columnwidth]{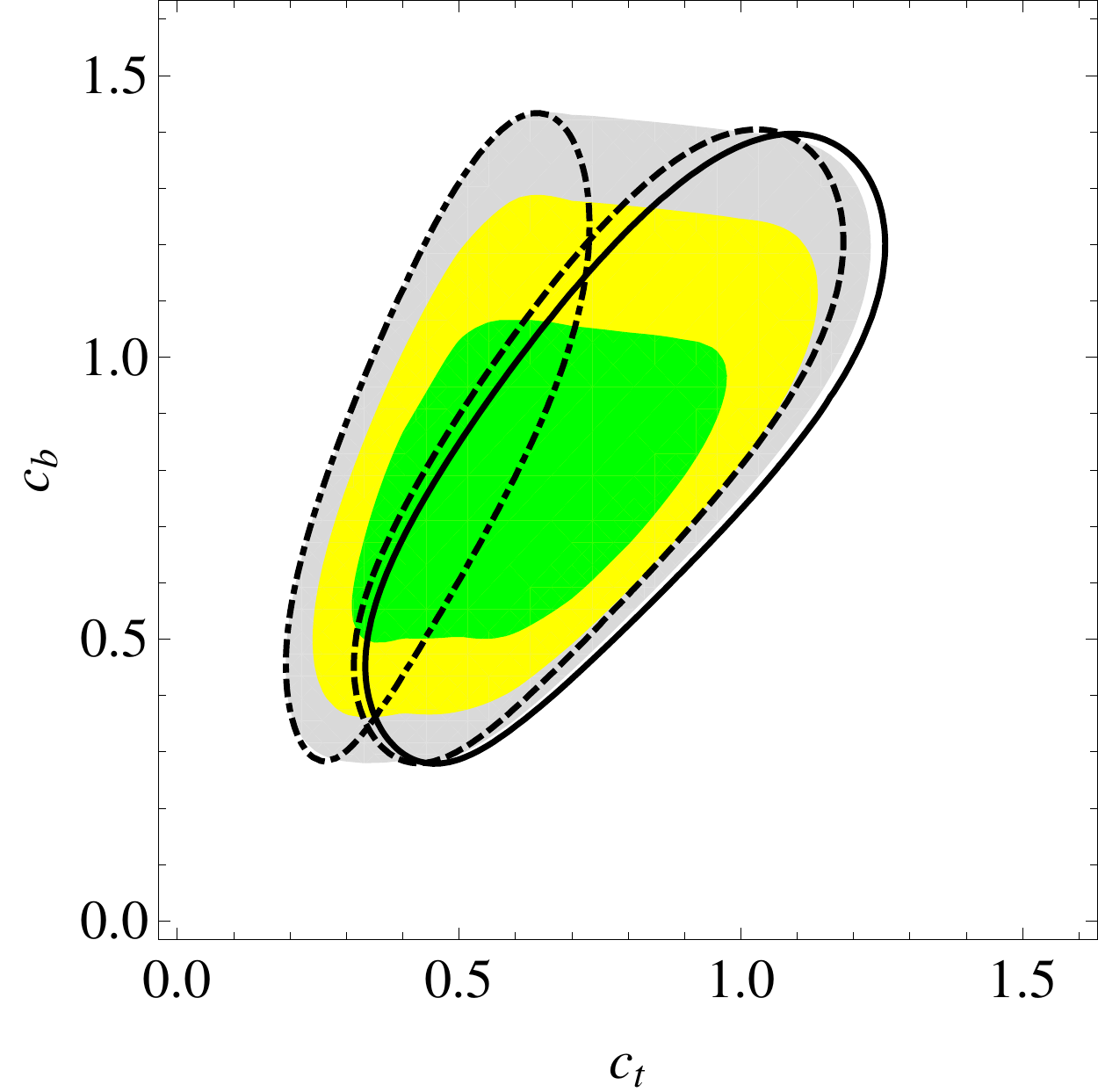}
\end{tabular}
\end{center}
\caption{\emph{99\% C.L. contours from a global fit to the parameters $c_b$ and $c_t$, taking into account the loop effects due to a stop quark with $m_{\tilde{t}}=160,500, \GeV$ (dot-dashed,dashed); in the solid line this effect is not taken into account, while in the colored contours stops (color coding as in Fig.~\ref{stop}) are treated as a nuisance and their contribution is marginalized  assuming  $m_{\tilde{t}}>160 \GeV$.}}
\label{fig:lightstops}
\end{figure}

So far we have studied modifications of the direct couplings between Higgs and fermions, restricting our attention to the 2HDM structure of the Higgs sector. When comparing with data, however, the presence of light sparticles  can introduce additional nuisances, as they contribute via loop-effects to the $hgg$ and $h\gamma\gamma$ effective vertices. Naturalness suggests that only the partners of third family fermions be light (other bounds on natural SUSY have been studied in Refs.~\cite{Antusch:2012gv,Arvanitaki:2011ck,ArkaniHamed:2012kq}); these have also the strongest couplings to the Higgs sector and have potentially a bigger impact than other sparticles. While staus and sbottoms have a negligible effect, light stops can change the analysis considerably \cite{Blum:2012ii}. We show this in Fig.~\ref{fig:lightstops}, where we compare 99\%C.L. contours, assuming that a stop has been found, with mass $m_{\tilde{t}}=160,500\GeV$ (dotted, dashed), with the contours without taking this effect into account (solid)\footnote{Recall that in our plots of the MSSM Figs.~\ref{stop}-\ref{stop2}, since heavy stops are needed to increase the Higgs mass, we have assumed heavy stops in the fit too, with no seizable loop contributions to $hgg$ and $h\gamma\gamma$; in Figs.~\ref{dterm}-\ref{nmssmix}, on the other hand, where a natural spectrum is allowed thanks to the D-term/F-term contributions to the Higgs mass, we assumed $m_{\tilde{t}}=500\GeV$.}.
Since $c_t$ itself affects Higgs physics mostly through a modification of the Higgs-gluon effective vertex\footnote{$c_t$ enters also directly through a contribution of a few percent of the $\bar t t h$ associated production channel to the total production crossection and through the exclusive  $p p\to h\bar t t\to\bar b b\bar t t$ channel, which is however badly measured at present.}, the leading effect of light stops, which themselves affect the $hgg$ effective coupling, results in a shift along the direction of $c_t$.
As it can be seen, stops heavier than about 500 GeV have negligible influence on the fit. Nevertheless, in Fig.~\ref{fig:lightstops} we also show the global fit treating the stop contribution as nuisance and marginalizing over it: this is useful to take into account the possibility that a very light stop lies in a region inaccessible to direct searches. 

The mass of charginos is also directly related to the EW scale if the chargino is mostly Higgsino: then , In principle light charginos introduce an additional unknown through their contribution to the $h\gamma\gamma$ coupling. However, for this to have any impact,  small $\tan\beta\lesssim 5$ \cite{Blum:2012ii}, very light charginos $m_{\chi^\pm}\ll 250 \GeV$~\cite{Djouadi:1996pb} and large wino-chargino mixing (which is typically suppressed by inverse powers of the wino mass $m_W^2/M_2^2$ \cite{Espinosa:2012in}  ) are necessary. We consider this a peculiar, rather than representative, point in parameter space and we assume these effects to be small in our analysis.

\section{Conclusions and Outlook}

In the MSSM, the tree-level Higgs mass is too small to account for the observed value of $\approx 125$ GeV. We have shown, using a simple analytical method based on an expansion in inverse powers of the heavy Higgs mass, how the physics that contributes to increase the light Higgs mass, also modifies the couplings of the lightest CP-even Higgs with fermions and gauge bosons. In the simplest examples (MSSM with no mixing between top squarks, MSSM with extra non-decoupling D-terms/F-terms) this connection provides distinctive predictions for the Higgs couplings, which allow us to extract bounds on the parameters $m_A$ and $\tan\beta$, competitive with bounds from direct searches~\cite{CMSHtautau}. Deviations in the couplings $hZZ$ and $hWW$ are expected to be small in 2HDMs and we could use the intuitive $c_b,c_t$ plane to show our results. In this way we could extract bounds on the heavy Higgs mass $m_H$, depending on the model and on the size of $\tan\beta$, as we summarized in Fig.~\ref{fig:excl}.

Models that include extra gauge singlets that mix with the Higgs sector, can in principle be studied in a similar way. In the most popular realizations, such as the NMSSM, however, the large number of parameters of the model weakens the Higgs mass/coupling connection and predictability is compromised. Yet, theoretical consistency of the model, based for instance on the requirement of perturbativity, can strongly constrain the size of the expected effects.

As long as the uncertainty in  measuring Higgs couplings is dominated by statistical errors, more data will lead to better measurments. In the long term, with an integrated luminosity  $\sim300 \textrm{fb}^{-1}$, the sensitivity to the Higgs couplings to bottom/top quarks is expected to reach 15\% \cite{Peskin:2012we,Linssen:2012hp} and, as we show in the right panel of Fig.~\ref{fig:excl}, some deviations from the SM are expected if $m_{H,A}\lesssim 400 \GeV$. At the same time, direct searches would have probed a much larger region of parameter space, but the bounds from Higgs couplings will remain competitive in the intermediate $\tan\beta$ region (better results can be achieved by considering ratios of couplings \cite{Djouadi:2012rh}).

Let us conclude with a comment regarding bounds from flavour physics. The cross-section for  $B_s \to \mu^+ \mu^-$ processes is proportional to  $\tan^6 \beta$ and therefore this measurement practically excludes  the region $\tan\beta\gtrsim  30$ (depending on other parameters of the model \cite{Altmannshofer:2012ks}) while it has a relatively small impact for intermediate and small $\tan\beta$; in this regime the bounds discussed in this paper can be considered complementary.  Constraints from $b \to s \gamma$ can be more important \cite{Carena:2000uj}, but a fair comparison is difficult, as the amplitude for this process depends as much on the details of the sparticle sector as it depends on the parameters of the Higgs sector, which we consider here (in regions where the former are small, bounds on Type II 2HDMs exclude $m_{H^\pm} \lesssim 300$ GeV, independently of $\tan \beta$~Ê\cite{Branco:2011iw}). In any case, while the $b \to s \gamma$ bounds are competitive with the Higgs coupling bounds at present\footnote{An interesting contribution to $b\to s\gamma$, which we have neglected throughout this work, comes from loop-effects involving squark-charginos or sbottom-gluino,  which do not decouple when the mass of the  superpartners is large and induce a coupling of $H_2$ to down-type quarks, $y_b( H_1^0 b\bar{b}+\epsilon_b H_2^0 b\bar{b})$ that can be enhanced at large $\tan\beta$ (which is interesting  for the MSSM and for the D-term case of section~\ref{sec:DTerms}). This can strengthen (weaken) the bounds from $b\to s \gamma$  for negative (positive)  $\epsilon_b$~\cite{Degrassi:2000qf}. At the same time, however, it can affect the $hbb$ coupling~\cite{Blum:2012ii},
\beq
\Delta c_b\approx -\epsilon_b \tan \beta \frac{m_h^2}{m_H^2},
\eeq
which, for the MSSM or for the D-terms of \eq{eol}, goes also towards strengthening (weakening) our bounds. Thus in the MSSM (also with additional D-terms)  stronger constraints from Higgs coupling data  are correlated with stronger $b \to s \gamma$  constraints and vice-versa~\cite{Gupta:2012mi}.}, the latter are expected to become stronger as the integrated luminosity increases.

\vspace{1cm}
\noindent
{\bf Note Added:} While this work was being finalized (see \cite{talkMarc}), Ref.~\cite{D'Agnolo:2012mj} appeared, which also considers modifications of Higgs couplings  in SUSY models with additional, non-decouplings F-terms or D-terms.

\section*{Acknowledgements}
We are grateful to A.~Pomarol for many discussions and comments that have helped shaping this article. We are also thankful to J.R.~Espinosa and J.~Virto for stimulating discussions. FR acknowledges support from the Swiss National Science Foundation, under the Ambizione grant PZ00P2\_136932.

\section*{APPENDIX I: Details of experimental fit}
To perform the statistical analysis of the data we take the following reescaled couplings of the higgs to the SM particles

\begin{equation}\label{equation1}
c_t\equiv \frac{y_t}{y_t^{SM}},\,\,c_b = c_\tau \equiv\frac{y_b}{y_b^{SM}},\,\,c_V\equiv\frac{g_{hVV}}{g_{hVV}^{SM}},
\end{equation}

but fixing $c_V=1$ the reason of which is explained before in the text.

We also take into account the stop loop effects which appear in the $gg \rightarrow h$, $h \rightarrow \gamma \gamma$, $h \rightarrow gg$ production and decay modes. The main contributions to these effects \cite{Dawson:1996xz,Spira:1995rr,Djouadi:1998az} are given by the following expressions:

\begin{equation}\label{equation1}
\frac{\sigma(gg \rightarrow h)}{\sigma(gg \rightarrow h)_{SM}} \approx \frac{\Gamma(h \rightarrow gg)}{\Gamma(h \rightarrow gg)_{SM}} \approx \left| \frac{A^{gg}_t+A^{gg}_{\tilde{t}_1}+A^{gg}_{\tilde{t}_2}}{A^{gg}_{t,SM}} \right|^2
\end{equation}

\begin{equation} \approx c_t^2 \left| 1+ \frac{m_t^2}{4 m_{\tilde{t}_1}}+\frac{m_t^2}{4 m_{\tilde{t}_2}}-\frac{m_t^2 X_t^2  }{4 m_{\tilde{t}_1} m_{\tilde{t}_2}} \right|^2
\end{equation}

\begin{equation}\label{equation1}
\frac{\Gamma(h \rightarrow \gamma \gamma)}{\Gamma(h \rightarrow \gamma \gamma)_{SM}} \approx \left| \frac{A^{\gamma \gamma}_w+A^{\gamma \gamma}_t+A^{\gamma \gamma}_{\tilde{t}_1}+A^{\gamma \gamma}_{\tilde{t}_2}}{A^{\gamma \gamma}_{w,SM}+A^{\gamma \gamma}_{t,SM}} \right|^2
\end{equation}

\begin{equation}
\approx \left| 1.28 \, a-0.28 \, c_t \left( \frac{m_t^2}{4 m_{\tilde{t}_1}}+\frac{m_t^2}{4 m_{\tilde{t}_2}}-\frac{m_t^2 X_t^2 }{4 m_{\tilde{t}_1} m_{\tilde{t}_2}}\right)  \right|^2
\end{equation}

The statistical analysis is performed using the latest signal strenght data given by Tevatron and ATLAS, and the one given by CMS at ICHEP. We didn't take the latest CMS data due to the fact that only a combination of 7 and 8 TeV is given in the signal strenghts at a higgs mass different than the one of ICHEP which doesn't allow us to extract them separatedly \footnote{An alternative approach can be found in \cite{azatovgall}}.

The signal strengths are assumed to follow a Gaussian distribution and we fit the data by minimizing a $\chi^2$ with the theoretical prediction for the signal strenght:

\begin{equation}
\mu^i=\frac{\sum_p\sigma_p(a,c_t,c_b,c_\tau)\zeta_p^i}{\sum_p\sigma_p^{SM}\zeta_p^i}\frac{BR_i(a,c_t,c_b,c_\tau)}{BR_i^{SM}},
\label{signalstrength}
\end{equation}

In the few cases where the given signal strenght errors are not symmetric we symmetrize them in quadrature. Statistical and theoretical errors are summed in quadrature without taking into account possible correlations, this approach is reasonable since at the moment the effect of this correlations is still small and can be neglected. In the other hand when comparing our fits using ICHEP data for both CMS and ATLAS we find an agreement of better than $\% 10$ between our figure and the one provided by them (we find this agreement to become better depending on the assumed cuts in the channels where they are not completely specified).

The data used can be found in table \ref{tableChannels} where all the channels taken into account are specified. In this table we see that for each channel a particular set of cuts is defined. These refer to the values of $\zeta_p^i$ found in equation \ref{signalstrength} which are the cuts for each higgs production mode: gluon fusion (G), vector boson fusion (VBF), associated production with a vector boson (A) and associated $t \bar{t}$ (tth). Expanded this can be seen as:

\begin{eqnarray}
\frac{\sum_p\sigma_p\zeta_p^i}{\sum_p\sigma_p^{SM}\zeta_p^i}=&\frac{c_t^2( \, \sigma_{G}\zeta_{G}^i + \,\sigma_{ttH}\zeta_{ttH}^i)+a^2( \sigma_{VBF}\zeta_{VBF}^i+ \sigma_{WH}\zeta_{WH}^i+\sigma_{ZH}\zeta_{ZH}^i)}
{\sigma_{G}\zeta_{G}^i+\sigma_{VBF}\zeta_{VBF}^i+\sigma_{WH}\zeta_{WH}^i+\sigma_{ZH}\zeta_{ZH}^i+\sigma_{ttH}\zeta_{ttH}^i},
\nonumber
\end{eqnarray}
where the cut efficiencies $\zeta_p^i$ for each production mode $p$ corresponding to each channel $i$ are reported below in table \ref{TableCutsI}.

\begin{table}[!h]
\caption{\emph{CMS, ATLAS  and  Tevatron data for the most sensitive channels. The cuts are classified as inclusive (I), associated production (A), vector boson fusion (VBF) or else ($\gamma\gamma_X$), see Appendix for details.
$\hat{\mu}^{1.96,7,8}$ denote the best fits for the 1.96 TeV Tevatron, and the 7,8 TeV  LHC data.}}

\begin{center}
\centering
\begin{tabular}{c c}
\begin{tabular}[t]{|c|c|c|c|}
\hline
$\begin{array}{c}\textrm{CMS}\\125\GeV\end{array}$  & Cuts & $\hat{\mu}^7$ & $\hat{\mu}^8$   \\ [2 pt]
\hline
 $\gamma \gamma_0$ \cite{CMSPhotons} & $\gamma\gamma_X$ & $3.1^{+1.9}_{-1.8}$  & $1.5^{+1.3}_{-1.3}$\\ [2 pt]
 $\gamma \gamma_1$ \cite{CMSPhotons}& $\gamma\gamma_X$ & $0.6^{+1.0}_{-0.9}$ & $1.5^{+1.1}_{-1.1}$ \\ [2 pt]
 $\gamma \gamma_2$ \cite{CMSPhotons} & $\gamma\gamma_X$& $ 0.7^{+1.2}_{-1.2} $& $1.0^{+1.2}_{-1.2}$\\ [2 pt]
 $\gamma \gamma_3$ \cite{CMSPhotons} & $\gamma\gamma_X$ & $1.5^{+1.6}_{-1.6}$ & $3.8^{+1.8}_{-1.8}$\\ [2 pt]
 $\gamma \gamma_{jj}$ \cite{CMSPhotons} & $\gamma\gamma_X$
   & $4.2^{+2}_{-2}$ & $\begin{array}{lr}L:&-0.6^{+2.0}_{-2.0}\\T:&1.3^{+1.6}_{-1.6}\end{array}$ \\ [2 pt]
 $\tau \tau_{0/1 j}$  \cite{CMSAllCh} & I & $1.0^{+1.5}_{-1.4}$ & $2.1^{+1.5}_{-1.6}$\\
 $\tau \tau_{VBF}$  \cite{CMSAllCh}  &  VBF  & $-1.8^{+1.4}_{-1.2}$ & $-1.8^{+1.4}_{-1.3}$\\
 $\tau \tau_{VH}$  \cite{CMSAllCh} &  A & $0.6^{+4.2}_{-3.1}$ & -\\
$ b b_{VH}$  \cite{CMSAllCh}& A & $0.6^{+1.3}_{-1.2}$& $0.4^{+1.2}_{-0.9}$ \\
$ b b_{ttH}$ \cite{CMSAllCh} & ttH & $-0.8^{+2.1}_{-1.8}$& - \\
$WW_{0j}$   \cite{ATLAS} & I & $0.1^{+0.6}_{-0.6}$ & $1.3^{+0.8}_{-0.6}$  \\
$WW_{1j}$  \cite{ATLAS}  & I & $1.7^{+1.2}_{-1.0}$ & $0.0^{+0.8}_{-0.8}$  \\
$WW_{2j}$  \cite{ATLAS} & VBF & $0.0^{+1.3}_{-1.3}$ & $1.3^{+1.7}_{-1.3}$ \\
$ZZ$ \cite{CMSAllCh} & I & $0.6^{+0.8}_{-0.5}$& $0.8^{+0.7}_{-0.5}$ \\
\hline
\hline
$\begin{array}{c}\textrm{CDF/D0}\\125\GeV\end{array}$   & Cuts & $\hat{\mu}^{1.96}$ & - \\
\hline
$\gamma \gamma$  \cite{:2012cn}& I & $3.6^{+3.0}_{-2.5}$& - \\
$bb$ \cite{:2012cn}& A & $2.0^{+0.7}_{-0.6}$  & - \\
$WW$ \cite{:2012cn}& I & $0.3^{+1.2}_{-0.3}$& - \\
\hline
\end{tabular} &
\begin{tabular}[t]{|c|c|c|c|}
\hline
$\begin{array}{c}\textrm{ATLAS}\\126.5\GeV\end{array}$  & Cuts & $\hat{\mu}^7$  & $\hat{\mu}^{8}$     \\
\hline
$ \gamma \gamma_{UnCeLPTt}$ \cite{ATLASPhotons,ATLASPhotons8}& $\gamma\gamma_X$ & $0.5^{+1.4}_{-1.4}$& $1.0^{+0.9}_{-0.9}$\\
$ \gamma \gamma_{UnCeHPTt}$ \cite{ATLASPhotons,ATLASPhotons8}& $\gamma\gamma_X$ & $0.2^{+2.0}_{-1.9}$& $0.3^{+1.7}_{-1.7}$\\
$ \gamma \gamma_{UnReLPTt}$ \cite{ATLASPhotons,ATLASPhotons8}& $\gamma\gamma_X$ & $2.5^{+1.7}_{-1.7}$& $2.9^{+1.2}_{-1.2}$\\
$ \gamma \gamma_{UnReHPTt}$ \cite{ATLASPhotons,ATLASPhotons8}& $\gamma\gamma_X$ & $10.4^{+3.7}_{-3.7}$& $1.8^{+1.4}_{-1.4}$\\
$ \gamma \gamma_{CoCeLPTt}$ \cite{ATLASPhotons,ATLASPhotons8}& $\gamma\gamma_X$ & $6.1^{+2.7}_{-2.7}$& $1.5^{+1.3}_{-1.3}$\\
$ \gamma \gamma_{CoCeHPTt}$ \cite{ATLASPhotons,ATLASPhotons8}& $\gamma\gamma_X$ & $-4.4^{+1.8}_{-1.8}$& $1.0^{+1.6}_{-1.6}$\\
$ \gamma \gamma_{CoReLPTt}$ \cite{ATLASPhotons,ATLASPhotons8}& $\gamma\gamma_X$ & $2.7^{+2.0}_{-2.0}$& $2.3^{+1.2}_{-1.2}$\\
$ \gamma \gamma_{CoReHPTt}$ \cite{ATLASPhotons,ATLASPhotons8}& $\gamma\gamma_X$ & $-1.6^{+2.9}_{-2.9}$& $0.5^{+1.6}_{-1.6}$\\
$ \gamma \gamma_{CoTr}$ \cite{ATLASPhotons,ATLASPhotons8}& $\gamma\gamma_X$ & $0.3^{+3.6}_{-3.6}$& $2.0^{+2.2}_{-2.2}$\\
$ \gamma \gamma_{2j}$ \cite{ATLASPhotons,ATLASPhotons8}& $\gamma\gamma_X$ & $2.7^{+1.9}_{-1.9}$& $\begin{array}{lr}L:&3.6^{+2.1}_{-2.1}\\H:&2.0^{+1.1}_{-1.1}\end{array}$\\
$ \gamma \gamma_{LepTag}$ \cite{ATLASPhotons8}& $\gamma\gamma_X$ & - & $1.2^{+2.4}_{-2.4}$\\
 $\tau \tau$   \cite{ATLAS:2012tx019,ATLAS:2012tx019WW}& I & $0.3^{+1.7}_{-1.8}$ & $0.73^{+0.71}_{-0.71}*$ \\
$ b b$  \cite{ATLAS:2012tx019,ATLAS:2012tx019bb}& A & $-2.7^{+1.6}_{-1.6}$& $1.0^{+1.4}_{-1.4}$\\
$WW$  \cite{ATLAS:2012tx019,ATLAS:2012tx019WW}& I & $0.5^{+0.6}_{-0.6}$ & $1.4^{+0.5}_{-0.6}$ \\
$ZZ$  \cite{ATLAS:2012tx019,ATLAS}& I & $1.1^{+1.0}_{-0.7}$& $0.9^{+0.7}_{-0.7}$ \\
\hline
\end{tabular}
\end{tabular}
\end{center}
\label{tableChannels}
\end{table}%

\begin{table}[!h]
\caption{Cut efficiencies for production modes \cite{CMSPhotons,ATLASPhotons,ATLASPhotons8} of the channels of table~\ref{tableChannels}. Numbers in brackets give the efficiencies at 8 TeV, the others at 7 TeV  and the overall normalization in each line factorizes.
 }
\begin{center}
\begin{tabular}{|c|c|c|c|c|c|}
\hline
$i$   & $\zeta_{G}^i$ & $\zeta_{VBF}^i$ & $\zeta_{WH}^i$ &  $\zeta_{ZH}^i$ &$\zeta_{ttH}^i$ \\ \hline
 $\gamma\gamma_0$ & 0.28(0.45) & 1(1) &1.52(1.91) & 1.52(1.91) &2.37(4.00) \\ \hline
 $\gamma\gamma_1$ & 1.16(1.17) & 1(1) &1.36(1.43)&1.36(1.43)&2.24(2.00) \\ \hline
 $\gamma\gamma_2$ & 1.80(1.84) & 1(1) &1.36(1.07) &1.36(1.07) &0(0) \\ \hline
 $\gamma\gamma_3$ & 1.80(1.84) & 1(1) &1.36(1.43) &1.36(1.43) &0(0)\\ \hline
  $\gamma\gamma_{jj}$ & 0.029 & 1 &0.019&0.019& 0 \\ \hline
  $\gamma\gamma_{jj}$ (T) &(0.024) & (1) &(0)&(0)& (0) \\ \hline
  $\gamma\gamma_{jj}$(L) &(0.094) & (1) &(0.064)&(0.064)& (0) \\ \hline
$\gamma\gamma_{UnCeLPTt}$ & 1.85 (1.78) & 1 (1) & 0.97 (0.77) & 0.99 (0.87) & 0.74 (0.58) \\ \hline 
$\gamma\gamma_{UnCeHPTt}$ &  0.34 (0.41) & 1 (1) & 1.36 (0.59) & 1.44 (0.77) & 2.27 (1.37) \\ \hline 
$\gamma\gamma_{UnReLPTt}$ & 1.90 (1.79) & 1 (1) & 1.11 (0.93) & 1.12 (1.06) & 0.76 (0.58) \\ \hline  
$\gamma\gamma_{UnReHPTt}$ & 0.32 (0.41) & 1 (1) & 1.45 (0.69) & 1.50 (0.89) & 1.66 (0.97) \\ \hline  
$\gamma\gamma_{CoCeLPTt}$ & 1.85 (1.78) & 1 (1) & 1.03 (0.77) & 0.99 (0.87) & 0.74 (0.58) \\ \hline  
$\gamma\gamma_{CoCeHPTt}$ & 0.35  (0.43) & 1 (1) & 1.41 (0.65) & 1.48 (0.78) & 2.43 (1.44) \\ \hline  
$\gamma\gamma_{CoReLPTt}$ & 1.95 (1.83) & 1 (1) & 1.14 (0.95) & 1.15 (1.09) & 0.78 (0.60) \\ \hline  
$\gamma\gamma_{CoReHPTt}$ & 0.33 (0.40) & 1 (1) & 1.49 (0.75) & 1.46 (0.91) & 1.67 (1.05) \\ \hline  
$\gamma\gamma_{CoTr}$ & 1.37 (1.30) & 1 (1) & 1.37 (0.95) & 1.30 (1.09) & 0.86 (0.66) \\ \hline  
$\gamma\gamma_{2j}$ & 0.02  & 1  & 0.01  & 0.01  & 0.02  \\ \hline  
$\gamma\gamma_{2j}$ (H) & (0.037)  & (1)  & (0.010)  & (0.012)  & (0.018)  \\ \hline  
$\gamma\gamma_{2j}$ (L) & (0.95)  & (1)  & (9.3)  & (9.6)  & (3.8)  \\ \hline  
$\gamma\gamma_{LepTag}$  & (0.65)  & (1)  & (359.4)  & (160.2)  & (551.4)  \\ \hline  
I & 1 & 1 & 1 & 1 & 1 \\ \hline
A & 0 & 0 & 1 & 1 & 0 \\ \hline
VBF & 0.029 & 1 &0.019&0.019& 0 \\ \hline
ttH & 0 & 0 & 0 & 0 & 1 \\ \hline
\end{tabular}
\end{center}
\label{TableCutsI}
\end{table}

\section*{APPENDIX II: Details of the exact theory computation}
\label{app2}

The most general two Higgs doublet model (2HDM) potential for the neutral components of the doublet is,
\bea
\Delta V& =& m_1^2 H_1^2+m_2^2H_2^2-(m_{12}^2 H_1 H_2+c.c.)+\frac{\lambda_1}{2} |H_1|^4+ \frac{\lambda_2}{2} |H_2|^4+ \lambda_3 |H_1|^2|H_2|^2+  \lambda_4 (H_1H_2)^\dagger(H_1H_2) \nonumber\\&+&( \frac{\lambda_5} {2}|H_1H_2|^2 +\lambda_6  |H_1|^2 H_1H_2+\lambda_7  |H_2|^2 H_1H_2+c.c)
\eea
We have used the convention of Ref.~Ê\cite{Carena:1995bx, Haber:1993an}. We can rewrite this potential in the $h$-$H$ basis to obtain,
\bea
\delta&=&  \frac{\lambda_1}{2}  c_\beta^3 s_\beta+ \frac{\lambda_2}{2}  s_\beta^3 c_\beta+ \frac{1}{2}(\lambda_3+\lambda_4+\lambda_5) s_\beta c_\beta (s_\beta^2-c_\beta^2)  +\frac{ \lambda_6 }{2}(c_\beta^2(s_\beta^2-c_\beta^2)+2s_\beta^2c_\beta^2)+\frac{ \lambda_7 }{2}(s_\beta^2(s_\beta^2-c_\beta^2)-2s_\beta^2c_\beta^2)\nonumber\\
\delta_\lambda &=& \frac{\lambda_1}{8}  c_\beta^4 + \frac{\lambda_2}{8}  s_\beta^4 + \frac{1}{4}(\lambda_3+\lambda_4+\lambda_5) s^2_\beta c^2_\beta   +\frac{ \lambda_6 }{2}(c_\beta^3 s_\beta)+\frac{ \lambda_7 }{2}(s_\beta^3 c_\beta)
\eea
We will now give the values of $\lambda_1$-$\lambda_5$ in the different models we have considered.  In the MSSM we have,
\beq
\lambda_1 = \frac{m_Z^2}{2 v^2}~~~~~~~~~
\lambda_2 = \frac{m_Z^2}{2 v^2}~~~~~~~~~
\lambda_3 =- \frac{m_Z^2}{2 v^2}
\eeq
At the one loop level we get the following additional contributions to the effective potential from top squark loops,
\bea
 \Delta  \lambda_2 &=& \frac{y_t ^4}{32 \pi^2}(6 (2 + c_{21} l_s) l_s + (1 + c_{21} l_s) x_t a_t  y_t^4 (12 -
      x_t a_t))\nonumber\\
 \Delta  \lambda_5 &=&- \frac{y_t ^4}{32 \pi^2}(1 + c_{11} l_s) (\tilde{\mu})^2 x_t^2 y_t^4 \nonumber\\
 \Delta  \lambda_7 &=&-\tilde{\mu}  \frac{y_t ^4}{32 \pi^2}x_t  (6 - x_t a_t) (1 + c_{31} l_s)\nonumber\\
\eea
Here,
\bea
l_s&=& \log[M_s^2/m_t^2],~~y_t=m_t(m_t)/(v \sin \beta),\nonumber\\
x_t&=&(A_t- \mu \cot \beta)/M_s,~~\tilde{\mu}=\mu/M_s,~~Êa_t=A_t/M_s\nonumber\\
c_{11} &=& \frac{1}{32 \pi^2} (12 y_t^2 - 32 g_3^2(m_t)) \nonumber\\
c_{21} &=& \frac{1}{32 \pi^2} (6 y_t^2 - 32 g_3^2(m_t))\nonumber\\
c_{31} &=&- \frac{1}{32 \pi^2} (9 y_t^2 - 32 g_3^2(m_t))
\eea
For the BMSSM we get,
\beq
\Delta  \lambda_5 =2 \epsilon_1~~~~~~~~~
\Delta \lambda_6 =2 \epsilon_2~~~~~~~~~
 \Delta \lambda_7 = 2 \epsilon_2
\eeq
For the D-term extension   we get,
\beq
\Delta  \lambda_1 = 2\kappa~~~~~~~~~
\Delta \lambda_2 =2\kappa~~~~~~~~~
 \Delta \lambda_3 = - 2\kappa
\eeq
Finally for the NMSSM with no doublet singlet mixing we get,
\beq
\Delta \lambda_4 =| \lambda_S|^2
\eeq
The case of NMSSM with doublet singlet mixing has been dealt with in great detail in Sec.~Ê\ref{withmix}. For F-terms from triplets we get,
\beq
\Delta  \lambda_1 =2|\lambda_T|^2~~~~~~~~~
\Delta \lambda_2 =2|\lambda_{\bar T}|^2~~~~~~~~~
 \eeq
We can now write the mass matrix elements of the CP-even sector in terms of these couplings,
\bea
{\cal M}_{12}&=&2 v^2[(\lambda_3 + \lambda_4)s_\beta c_\beta + \lambda_6 c_\beta^2+ \lambda_7 s_\beta^2]-m_A^2 \s \c\\
{\cal M}_{12}&=&2 v^2[\lambda_1 c^2_\beta +2  \lambda_6 c_\beta s_\beta+ \lambda_5 s_\beta^2]+ m_A^2 \s^2\\
{\cal M}_{22}&=&2 v^2[\lambda_2 s^2_\beta +2  \lambda_7 c_\beta s_\beta+ \lambda_5 c_\beta^2]+ m_A^2 \c^2\
\eea
where we have used~Ê\cite{Haber:1993an},
\beq
m_{12}^2= \s \c(m_A^2 + 2 \lambda_5 v^2+ \lambda_6 t^{-1}_\beta v^2+\lambda_7 t_\beta v^2).
\eeq
Once we know the CP-even matrix elements we can easily  find the exact coupling deviations. First we demand that the light Higgs mass,
 \beq
m_h^2=\frac{1}{2}\left({\cal M}_{11}+{\cal M}_{22} - \sqrt{({\cal M}_{22}-{\cal M}_{11})^2- 4 {\cal M}^2_{12}}\right)
\label{hmass}
\eeq
is equal to 125 GeV by choosing an appropriate value of the stop mass in the MSSM, an appropriate value of $g_X$ for the D-term extension and an appropriate value of $\lambda$ for the NMSSM. Now we can compute   $c_b=-\sin\alpha/\cos\beta$ and $c_t=\cos\alpha/\sin\beta$, where $\alpha$ is extracted from
\beq
\tan 2 \alpha= \frac{2 {\cal M}_{12}}{{\cal M}_{11}-{\cal M}_{22}}.
\label{alpha}
\eeq


\end{document}